\pdfoutput=1
\documentclass[usenatbib]{mn2e}
\usepackage{apjfonts}
\usepackage{amssymb}
\usepackage{amsmath}
\usepackage{ctable}
\usepackage{url}
\usepackage{fixltx2e} 

\newcommand{\be}{\begin{equation}}
\newcommand{\ee}{\end{equation}}

\newcommand{\etal}{et al.}

\newcommand{\msun}{M_{\sun}}

\newcommand{\paperone}{Paper {\small I}}
\newcommand{\papertwo}{Paper {\small II}}
\newcommand{\paperthree}{Paper {\small III}}

\newcommand{\movieurl}{{\scriptsize\url{http://www.tapir.caltech.edu/~phopkins/Site/Movies_cosmo.html}}}
\newcommand{\FIREurl}{\url{http://fire.northwestern.edu}}
\newcommand{\gizmourl}{\url{www.tapir.caltech.edu/~phopkins/Site/GIZMO.html}}

\newcommand\plotonesize[2]
 {\centering \leavevmode \includegraphics[width={#2\columnwidth}]{#1}}
\newcommand{\plotsidesize}[2]
 {\centering \leavevmode \includegraphics[width={#2\textwidth}]{#1}}
\newcommand{\acknowledgments}{\begin{small}\section*{Acknowledgments}\end{small}}
\newcommand\altaffilmark[1]{$^{#1}$}
\newcommand\altaffiltext[1]{$^{#1}$}
\title[Galaxies on FIRE: Feedback \&\ Inefficient Star Formation]{Galaxies on FIRE (Feedback In Realistic Environments): Stellar Feedback Explains Cosmologically Inefficient Star Formation
\vspace{-0.5cm}}

\vspace{-0.2cm}
\author[Hopkins \etal]{
\parbox[t]{\textwidth}{ 
Philip F.~Hopkins\thanks{E-mail:phopkins@caltech.edu}\altaffilmark{1,2},
Du\v{s}an Kere\v{s}\altaffilmark{3}, 
Jos{\'e} O{\~{n}}orbe\altaffilmark{4},
Claude-Andr{\'e} Faucher-Gigu{\`e}re\altaffilmark{2,5}, 
Eliot Quataert\altaffilmark{2}, 
Norman Murray\altaffilmark{6,7}, \&\ 
James S.~Bullock\altaffilmark{4}
} 
\vspace*{6pt} \\
\altaffiltext{1}{TAPIR, Mailcode 350-17, California Institute of Technology, Pasadena, CA 91125, USA} \\
\altaffiltext{2}{Department of Astronomy and Theoretical Astrophysics Center, University of California Berkeley, Berkeley, CA 94720} \\
\altaffiltext{3}{Department of Physics, Center for Astrophysics and Space Science, University of California at San Diego, 9500 Gilman Drive, La Jolla, CA 92093} \\ 
\altaffiltext{4}{Department of Physics and Astronomy, University of California at Irvine, Irvine, CA 92697, USA} \\ 
\altaffiltext{5}{Department of Physics and Astronomy and CIERA, Northwestern University, 2145 Sheridan Road, Evanston, IL 60208, USA} \\ 
\altaffiltext{6}{Canadian Institute for Theoretical Astrophysics, 
60 St.\ George Street, University of Toronto, ON M5S 3H8, Canada} \\
\altaffiltext{7}{Canada Research Chair in Astrophysics\vspace{-0.5cm}} \\
\vspace{-0.5cm}
}

\date{Submitted to MNRAS, November, 2013\vspace{-0.6cm}}
\begin{document}
\maketitle
\label{firstpage}

\vspace{-0.2cm}
\begin{abstract}
\vspace{-0.2cm}

We present a series of high-resolution cosmological simulations$^{\ref{foot:movie}}$ of galaxy formation to $z=0$, spanning halo masses $\sim10^{8}-10^{13}\,\msun$, and stellar masses $\sim10^{4}-10^{11}\,\msun$. Our simulations include fully explicit treatment of the multi-phase ISM \&\ stellar feedback. The stellar feedback inputs (energy, momentum, mass, and metal fluxes) are taken directly from stellar population models. These sources of feedback, with {\em zero} adjusted parameters, reproduce the observed relation between stellar and halo mass up to $M_{\rm halo}\sim10^{12}\,\msun$. We predict weak redshift evolution in the $M_{\ast}-M_{\rm halo}$ relation, consistent with current constraints to $z>6$. We find that the $M_{\ast}-M_{\rm halo}$ relation is insensitive to numerical details, but is sensitive to feedback physics. Simulations with only supernova feedback fail to reproduce observed stellar masses, particularly in dwarf and high-redshift galaxies: radiative feedback (photo-heating and radiation pressure) is necessary to destroy GMCs and enable efficient coupling of later supernovae to the gas. Star formation rates agree well with the observed Kennicutt relation at all redshifts. The galaxy-averaged Kennicutt relation is very different from the numerically imposed law for converting gas into stars, and is determined by self-regulation via stellar feedback. Feedback reduces star formation rates and produces reservoirs of gas that lead to rising late-time star formation histories, significantly different from halo accretion histories. Feedback also produces large short-timescale variability in galactic SFRs, especially in dwarfs. These properties are not captured by common ``sub-grid'' wind models.

\end{abstract}

\begin{keywords}
galaxies: formation --- galaxies: evolution --- galaxies: active --- 
stars: formation --- cosmology: theory\vspace{-0.5cm}
\end{keywords}

\vspace{-1.1cm}
\section{Introduction}
\label{sec:intro}


\begin{figure*}
    \centering
\begin{tabular}{cc}
    \includegraphics[width={0.487\textwidth}]{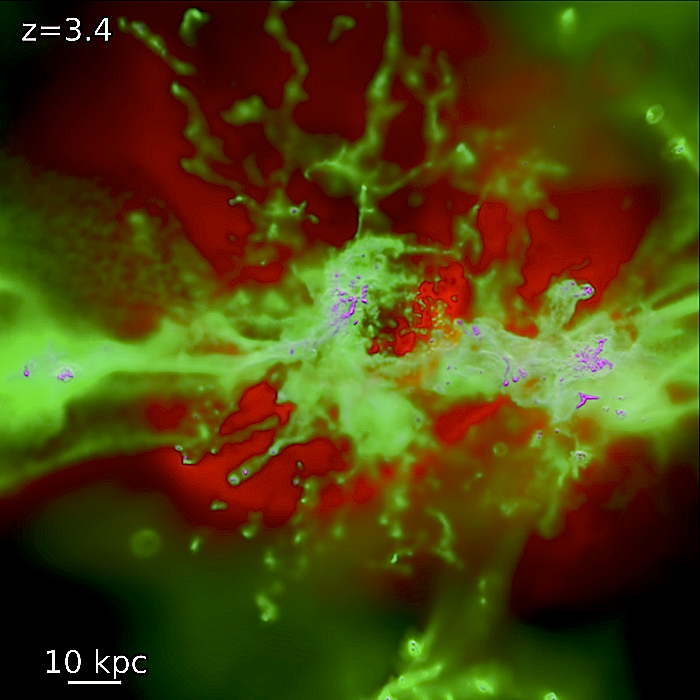} &
    \hspace{-0.4cm}
    \includegraphics[width={0.49\textwidth}]{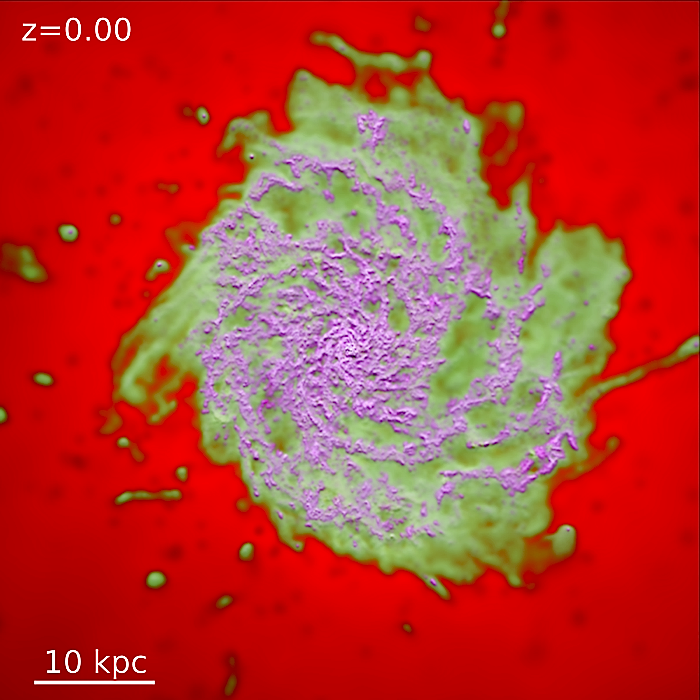} \\
\end{tabular}
    \vspace{-0.1cm}
    \caption{Gas in a representative simulation of a Milky Way-mass halo ({\bf m12i} in Table~\ref{tbl:sims}). Image shows the projected gas density, log-weighted ($\sim4\,$dex stretch). Magenta shows cold molecular/atomic gas ($T<1000\,$K). Green shows warm ionized gas ($10^{4}\lesssim T \lesssim 10^{5}\,$K). Red shows hot gas ($T\gtrsim 10^{6}\,$K).$^{\ref{foot:gas.coloring}}$ Each image shows a box centered on the main galaxy. {\em Left:} Box $200\,$kpc (physical) on a side at high redshift. The galaxy has undergone a violent starburst, leading to strong outflows of hot and warm gas that have blown away much of the surrounding IGM (even outside the galaxy). Note that the ``filamentary'' structure of cool gas in the IGM is clearly affected by the outflows. {\em Right:} Near present-day, with a $\sim50\,$kpc box. A more relaxed, well-ordered disk has formed, with molecular gas tracing spiral structure, and a halo enriched by diffuse hot outflows.
    \label{fig:demo.image.1}}
\end{figure*}


\begin{figure}
    \centering
    \plotonesize{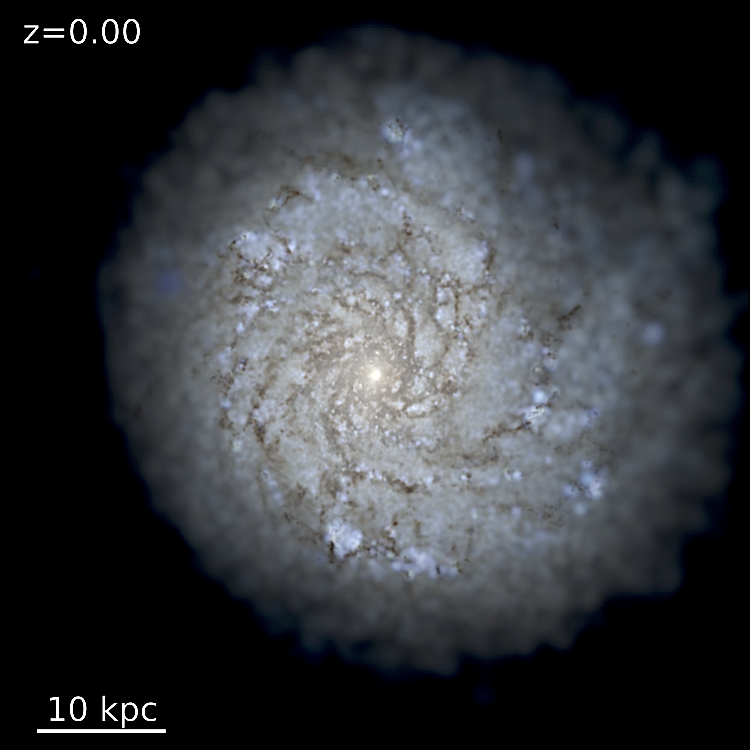}{1.01}
    \caption{Stars in the {\bf m12i} simulation at $z\sim0$, in a box $50\,$kpc on a side near present-time. Image is a mock $u/g/r$ composite. The disk is approximately face-on, and the spiral structure is visible. (The image uses {\small STARBURST99} to determine the SED of each star particle given its known age and metallicity, then ray-traces the line-of-sight flux following \citet{hopkins:lifetimes.letter}, attenuating with a MW-like reddening curve with constant dust-to-metals ratio for the abundances at each point.) 
    \label{fig:demo.image.2}}
\end{figure}

\begin{figure}
    \centering
\begin{tabular}{c}
    \includegraphics[width={0.9\columnwidth}]{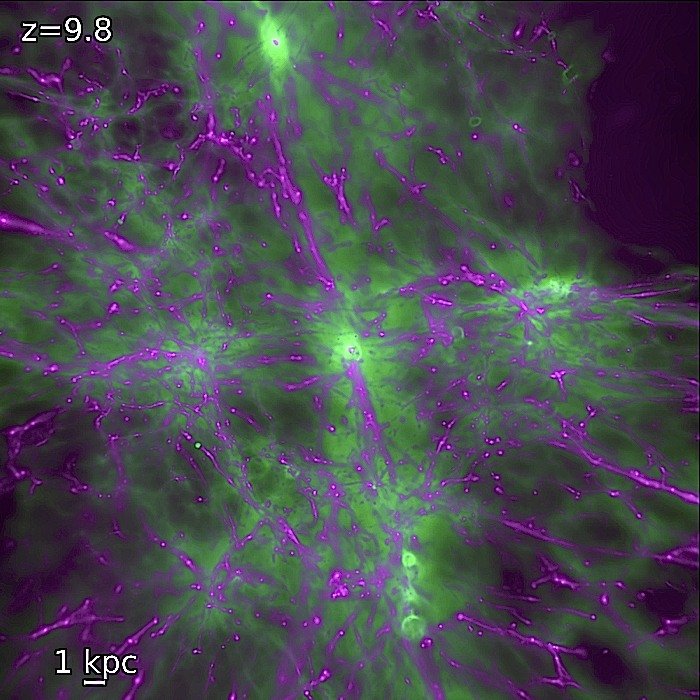} \\
    \includegraphics[width={0.9\columnwidth}]{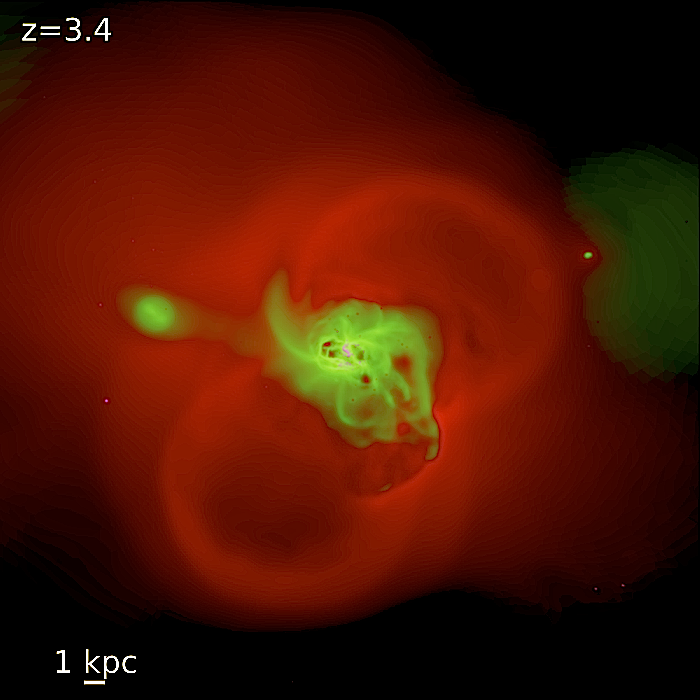} \\
\end{tabular}
    \caption{Gas, as Fig.~\ref{fig:demo.image.1}, for a dwarf galaxy ({\bf m10} in Table~\ref{tbl:sims}). {\em Top:} $40\,$kpc (physical) box, at high redshift. {\em Bottom:} $20\,$kpc box at intermediate redshift. Strong outflows are still present, though they are more spherical, because the galaxy halo is itself small and embedded {\em within} a much larger filament.
    \label{fig:demo.image.3}}
\end{figure}

It is well-known that feedback from stars is a critical, yet poorly-understood, component of galaxy formation. Within galaxies, star formation is observed to be inefficient in both an instantaneous and an integral sense.

Instantaneously, the Kennicutt-Schmidt (KS) relation implies gas consumption timescales of $\sim50$ dynamical times \citep{kennicutt98}, while the total fraction of GMC mass converted into stars is only a few percent \citep{zuckerman:1974.gmc.constraints,williams:1997.gmc.prop,evans:1999.sf.gmc.review,evans:2009.sf.efficiencies.lifetimes}. Without strong stellar feedback, however, gas inside galaxies cools efficiently and collapses on a dynamical time, predicting order-unity star formation efficiencies on all scales \citep{hopkins:rad.pressure.sf.fb,tasker:2011.photoion.heating.gmc.evol,bournaud:2010.grav.turbulence.lmc,dobbs:2011.why.gmcs.unbound,krumholz:2011.rhd.starcluster.sim,harper-clark:2011.gmc.sims}. 

In an integral sense, without strong stellar feedback, gas in cosmological models cools rapidly and inevitably turns into stars, predicting galaxies with far larger masses than are observed \citep[e.g.][and references therein]{katz:treesph,somerville99:sam,cole:durham.sam.initial,springel:lcdm.sfh,keres:fb.constraints.from.cosmo.sims}. Decreasing the instantaneous star formation efficiency does not eliminate this integral problem: the amount of baryons in real galactic disks is much lower than that predicted in models absent strong feedback \citep[essentially, the Universal baryon budget; see][]{white:1991.galform,keres:fb.constraints.from.cosmo.sims}. Constraints from intergalactic medium (IGM) enrichment require that many of those baryons must have entered galaxy halos and disks at some point to be enriched, before being expelled \citep{aguirre:2001.igm.metal.evol.sims,pettini:2003.igm.metal.evol,songaila:2005.igm.metal.evol,martin:2010.metal.enriched.regions}. Galactic super-winds with mass-loading $\dot{M}_{\rm wind}$ of many times the star formation rate (SFR) are therefore generally required to reproduce observed galaxy properties \citep[e.g.][]{oppenheimer:outflow.enrichment}. Such winds have been observed ubiquitously in local and high-redshift star-forming galaxies \citep{martin99:outflow.vs.m,martin06:outflow.extend.origin,heckman:superwind.abs.kinematics,newman:z2.clump.winds,sato:2009.ulirg.outflows,chen:2010.local.outflow.properties,steidel:2010.outflow.kinematics,coil:2011.postsb.winds}.

However, until recently, numerical simulations have been unable to produce winds with large-mass loading factors from an a priori model (let alone the correct scalings of wind mass-loading with galaxy mass or other properties), nor to simultaneously predict the instantaneous inefficiency of star formation within galaxies. This is particularly true of models which invoke only energetic feedback via supernovae (SNe), which is efficiently radiated in the dense gas where star formation actually occurs \citep[see e.g.][and references therein]{guo:2010.hod.constraints,powell:2010.sne.fb.weak.winds,brook:2010.low.ang.mom.outflows,nagamine:2010.dwarf.gal.cosmo.review,bournaud10}. More recent simulations, using higher resolution and invoking stronger feedback prescriptions, have seen strong winds, but have generally found it necessary to include simplified prescriptions for ``turning off cooling'' in the SNe-heated gas and/or include some adjustable parameters representing ``pre-SNe'' feedback \citep[see][]{governato:2010.dwarf.gal.form,maccio:2012.cuspcore.outflows,teyssier:2013.cuspcore.outflow,stinson:2013.new.early.stellar.fb.models,agertz:2013.new.stellar.fb.model}. This is physically motivated since feedback processes other than SNe -- protostellar jets, HII photoionization, stellar winds, and radiation pressure -- both occur and are critical in suppressing star formation in dense gas, as well as ``pre-processing'' gas prior to SNe explosions so that SNe occur at densities where thermal heating can have much larger effects \citep{evans:2009.sf.efficiencies.lifetimes,hopkins:rad.pressure.sf.fb,tasker:2011.photoion.heating.gmc.evol,lopez:2010.stellar.fb.30.dor,stinson:2013.new.early.stellar.fb.models,kannan:2013.early.fb.gives.good.highz.mgal.mhalo}.

And in fact, there have been many studies with enormously higher resolution (enough to evolve each star explicitly) and a full treatment of the radiation-magnetohydrodynamics and time dependence of these multiple feedback mechanisms. Because of computational limitations, however, these have necessarily been restricted to very small systems, either single molecular clouds/star clusters \citep[e.g.][]{krumholz:2007.rhd.protostar.modes,krumholz:2011.rhd.starcluster.sim,offner:2009.rhd.lowmass.stars,offner:2011.rad.protostellar.outflows,harper-clark:2011.gmc.sims,bate:2012.rmhd.sims}, or the ``first stars'' \citep[e.g.][]{wise:2012.rad.pressure.effects,pawlik:2013.rad.feedback.first.stars,muratov:2013.popIII.star.feedback}. But these studies, {\em without exception}, have found that the non-linear interaction of the feedback mechanisms above -- especially the dual roles of HII photoionization and radiation pressure in concert with SNe -- is absolutely critical to explain the generation of large local outflows, the self-regulation of star formation, and the shape of the stellar initial mass function. 

Despite these breakthroughs, given limited resolution and the complexity of the baryonic physics, many cosmological models have treated galactic wind generation and the inefficiency of star formation in a tuneable, ``sub-grid'' fashion. This is not to say that the models have not tremendously improved our understanding of galaxy formation! They have demonstrated that stellar feedback can {\em plausibly} lead to (globally) inefficient star formation, constrained the parameter space of allowed feedback models, made predictions for the critical role of outflows and recycling in enriching the IGM, provided possible baryonic solutions to apparent dark matter ``problems'' \citep[e.g.][]{pontzen:2011.cusp.flattening.by.sne}, demonstrated the need for ``early'' feedback from radiative mechanisms beyond SNe alone, and generally created the framework for our interpretation of observations. However, with wind models often relying on adjustable parameters, the integrated efficiency of star formation in galaxies is to some extent tuned ``by hand'' and the predictive power is inherently limited. This is particularly true for studies of gas in the circum-galactic medium (CGM), a current area of much observational progress -- measurements of the CGM are sensitive to the phase structure of the gas, which is not faithfully represented in models which simply ``turn off'' hydrodynamics or cooling, or mimic strong feedback via pure thermal energy injection or ``particle kicks'' \citep[see e.g.][for an explicit demonstration of this]{hummels:2013.cgm.vs.obs}.

Accurate treatment of star formation and galactic winds ultimately requires realistic treatment of the stellar feedback processes that maintain the multi-phase ISM. Motivated by this philosophy (and building on the studies with single-star resolution), in \citet{hopkins:rad.pressure.sf.fb} (\paperone) and \citet{hopkins:fb.ism.prop} (\papertwo), we developed a new set of numerical models to follow stellar feedback on scales from sub-GMC star-forming regions through galaxies. These simulations include the energy, momentum, mass, and metal fluxes from stellar radiation pressure, HII photo-ionization and photo-electric heating, SNe Types I \&\ II, and stellar winds (O-star and AGB). Critically, the feedback is directly tied to the young stars, with the energetics and time-dependence taken from stellar evolution models. In our previous work, we showed, in isolated galaxy simulations, that these mechanisms produce a quasi-steady ISM in which GMCs form and disperse rapidly, with phase structure, turbulence, and disk and GMC properties in good agreement with observations \citep[for various comparisons, see][]{narayanan:2012.mw.x.factor,hopkins:dense.gas.tracers,hopkins:clumpy.disk.evol,hopkins:2013.accretion.doesnt.drive.turbulence}. In \paperone, \citet{hopkins:virial.sf}, and \citet{hopkins:stellar.fb.mergers} we showed that this leads naturally to ``instantaneously'' inefficient SF (predicting the KS-law), regulated self-consistently by feedback and {\em independent} of the numerical prescription for star formation in very dense gas. In \citet{hopkins:stellar.fb.winds} (\paperthree) and \citet{hopkins:2013.merger.sb.fb.winds} we showed that the same feedback models reproduce the galactic winds invoked in previous semi-analytic and cosmological simulations, and that the {\em combination} of multiple feedback mechanisms is critical to produce massive, multi-phase galactic winds. 

However, our simulations have thus far been limited to idealized studies of isolated galaxies and galaxy mergers. These previous calculations thus cannot follow accretion from or interaction of outflows with the IGM, realistic galaxy merger histories, and many other important processes. In this paper, the first of a series, we present the FIRE (Feedback In Realistic Environments) simulations:\footnote{\label{foot:movie}Movies and summaries of key simulation properties are available at\\ \movieurl\\ and the {\small FIRE} project website:\\ \FIREurl} a suite of fully cosmological ``zoom-in'' simulations developed to study the role of feedback in galaxy formation. To test the models and understand feedback in a wide range of environments, we study a wide range in galaxy halo and stellar mass (as opposed to focusing just on MW-like systems), and follow evolution fully to $z=0$. Our suite of calculations includes several of the highest-resolution galaxy formation simulations that have been run to $z=0$. Our simulations utilize a significantly improved numerical implementation of SPH (which has resolved historical discrepancies with grid codes), as well as the full physical models for feedback and ISM physics introduced and tested in \paperone-\paperthree. Here, we explore the consequences of stellar feedback for the inefficiency of star formation, perhaps the most basic consequence of stellar feedback for galaxy formation. In companion papers, we will investigate the properties of outflows and their interactions with the IGM, the effect of those outflows on dark matter structure, the differences between numerical methods in treating feedback, the role of feedback in determining galaxy structure, and many other open questions.

In \S~\ref{sec:ICs}-\ref{sec:sims}, we describe our methodology. \S~\ref{sec:ICs} describes the initial conditions for the simulations; \S~\ref{sec:sims:physics} outlines the implementation of the key baryonic physics of cooling, star formation, and feedback (a much more detailed description is given in Appendix~\ref{sec:appendix:algorithms}); \S~\ref{sec:sims} briefly describes the improvements in the numerical method compared to past work (again, more details are in Appendix~\ref{sec:appendix:sims}). And in Appendix~\ref{sec:appendix:test.isolated} we test and compare these algorithms with higher-resolution simulations of isolated (non-cosmological) galaxies.

We describe our results in \S~\ref{sec:results}. We examine the predicted galaxy stellar masses (\S~\ref{sec:results:masses}), and how this depends on both numerical algorithms (\S~\ref{sec:results:numerics}) and feedback physics (\S~\ref{sec:results:feedback}), as well as how it compares to previous theoretical work (\S~\ref{sec:results:previous}). We show that the treatment of feedback physics overwhelmingly dominates these results, and discuss the distinct roles of multiple independent feedback mechanisms. We also explore the predictions for the Kennicutt-Schmidt relation (\S~\ref{sec:results:ks}), the shape of galaxy star formation histories (\S~\ref{sec:results:sfh}), the star formation ``main sequence'' (\S~\ref{sec:results:mainsequence}), and the ``burstiness'' of star formation (\S~\ref{sec:results:burstiness}). We summarize our important conclusions and discuss future work in \S~\ref{sec:discussion}.

\vspace{-0.5cm}
\section{Initial Conditions \&\ Galaxy Properties}
\label{sec:ICs}

The simulations presented here are a series of fully cosmological ``zoom-in'' simulations of galaxy formation; some images of the gas and stars in representative stages are shown in Figs.~\ref{fig:demo.image.1}-\ref{fig:demo.image.3}.\footnote{\label{foot:gas.coloring}Both gas and stellar images are true three-color volume renderings generated by ray-tracing lines of sight through the simulation (with every gas or star particle a source, respectively). For the stars, the physical luminosities and dust opacities in each band are used to generate the observed intensity map. For the gas, we construct synthetic ``bands'' where the particle emissivity is uniform if it falls within the temperature range specified, and zero otherwise, and the particle opacity is uniform across bands.} The technique is well-studied; briefly, a large cosmological box is simulated at low resolution to $z=0$, and then the mass within and around halos of interest at that time is identified, traced back to the starting redshift, and the Lagrangian region containing this mass is re-initialized at much higher resolution (with gas added) for the ultimate simulation \citep{porter:1985.cosmo.sim.zoom.outline,katz:1993.zoomin.technique}.

We consider a series of systems with different masses. Table~\ref{tbl:sims} describes the initial conditions. All simulations begin at redshifts $\sim100-125$, with fluctuations evolved using perturbation theory up to that point.\footnote{Initial conditions were generated with the {\small MUSIC} code \citep{hahn:2011.music.code.paper}, using second-order Lagrangian perturbation theory.} 

\begin{footnotesize}
\ctable[
  caption={{\normalsize Simulation Initial Conditions}\label{tbl:sims}},center,star
  ]{lccccccl}{
\tnote[ ]{Parameters describing the initial conditions for our simulations (units are physical): \\
{\bf (1)} Name: Simulation designation. \\
{\bf (2)} $M_{\rm halo}^{0}$: Approximate mass of the $z=0$ ``main'' halo (most massive halo in the high-resolution region). \\
{\bf (3)} $m_{b}$: Initial baryonic (gas and star) particle mass in the high-resolution region, in our highest-resolution simulations. \\ 
{\bf (4)} $\epsilon_{b}$: Minimum baryonic gravity/force softening (minimum SPH smoothing lengths are comparable or smaller). Recall, force softenings are adaptive (mass resolution is fixed); for more details see Appendix~\ref{sec:appendix:sims}.\\
{\bf (3)} $m_{dm}$: Dark matter particle mass in the high-resolution region, in our highest-resolution simulations. \\ 
{\bf (4)} $\epsilon_{dm}$: Minimum dark matter force softening (fixed in physical units at all redshifts). 
}
}{
\hline\hline
\multicolumn{1}{c}{Name} &
\multicolumn{1}{c}{$M_{\rm halo}^{0}$} &
\multicolumn{1}{c}{$m_{b}$} & 
\multicolumn{1}{c}{$\epsilon_{b}$} & 
\multicolumn{1}{c}{$m_{dm}$} & 
\multicolumn{1}{c}{$\epsilon_{dm}$} & 
\multicolumn{1}{c}{Merger} & 
\multicolumn{1}{c}{Notes} \\ 
\multicolumn{1}{c}{\ } &
\multicolumn{1}{c}{[$h^{-1}\,M_{\sun}$]} & 
\multicolumn{1}{c}{[$h^{-1}\,M_{\sun}$]} &
\multicolumn{1}{c}{[$h^{-1}$pc]} &
\multicolumn{1}{c}{[$h^{-1}\,M_{\sun}$]} &
\multicolumn{1}{c}{[$h^{-1}$pc]} &
\multicolumn{1}{c}{History} & 
\multicolumn{1}{c}{\ } \\ 
\hline
{\bf m09} & 1.9e9 & 1.8e2 & 1.0 & 8.93e2 & 20 & normal & isolated dwarf \\
{\bf m10} & 0.8e10 & 1.8e2 & 2.0 & 8.93e2 & 20 & normal & isolated dwarf \\
{\bf m11} &  1e11 &  5.0e3 & 5.0 & 2.46e4 & 50 & quiescent & -- \\ 
{\bf m12v} &  5e11 & 2.7e4 & 7.0 & 1.38e5 & 100 & violent & several $z<2$ mergers \\ 
{\bf m12q} & 1e12 & 5.0e3 & 7.0 & 1.97e5 & 100 & late merger & -- \\ 
{\bf m12i} & 1e12 & 3.5e4 & 10 & 1.97e5 & 100 & normal & large ($\sim10\,R_{\rm vir}$) box \\ 
{\bf m13} &  1e13 & 2.6e5 & 15 & 1.58e6 & 150 & normal & ``small group'' mass \\
\hline\hline\\
}
\end{footnotesize}

\begin{figure}
    \centering
    \plotonesize{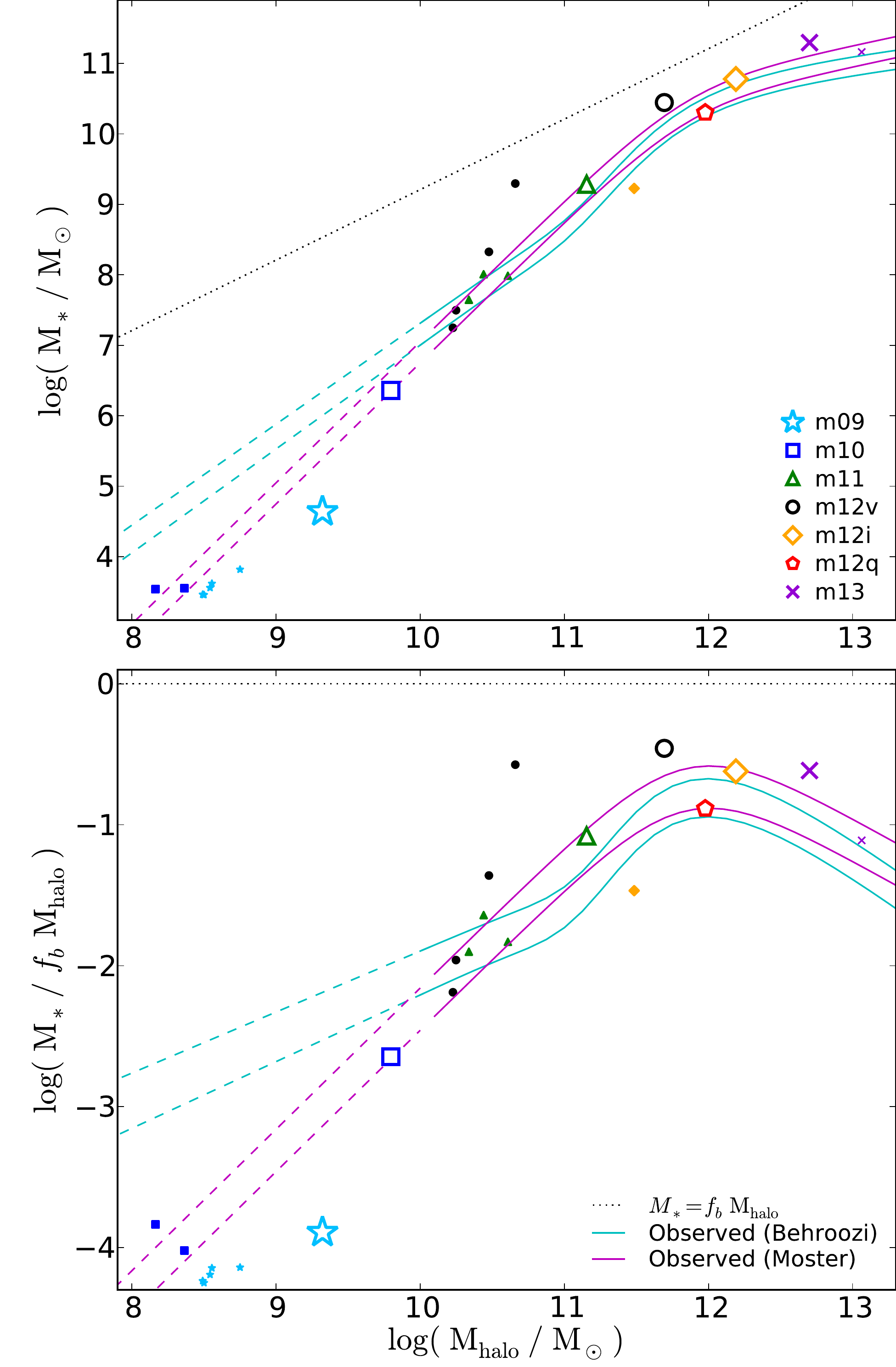}{0.99}
    \vspace{-0.2cm}
    \caption{Galaxy stellar mass-halo mass relation at $z=0$. {\em Top}: $M_{\ast}(M_{\rm halo})$. {\em Bottom:} $M_{\ast}$ relative to the Universal baryon budget of the halo ($f_{b}\,M_{\rm halo}$). Each simulation (points) from Table~\ref{tbl:sims} is shown; large point denotes the most massive halo in each box. We compare the relation if all baryons became stars ($M_{\ast}=f_{b}\,M_{\rm halo}$; dotted) and the observationally inferred relationship as determined in \citet[][magenta]{moster:2013.abundance.matching.sfhs} and \citet[][cyan]{behroozi:2012.abundance.matching.sfhs} (dashed lines denote extrapolation beyond the observed range).$^{\ref{foot:moster.behroozi.halo.defns}}$ The agreement with observations is excellent at $M_{\rm halo}\lesssim10^{13}\,\msun$, including dwarf though MW-mass galaxies. We stress that there are {\em zero} adjusted parameters here: stellar feedback, with known mechanisms taken from stellar population models, is sufficient to explain galaxy stellar masses at/below $\sim L_{\ast}$.
    \label{fig:mg.mh.z0}}
\end{figure}

\begin{figure*}
    \centering
    \plotsidesize{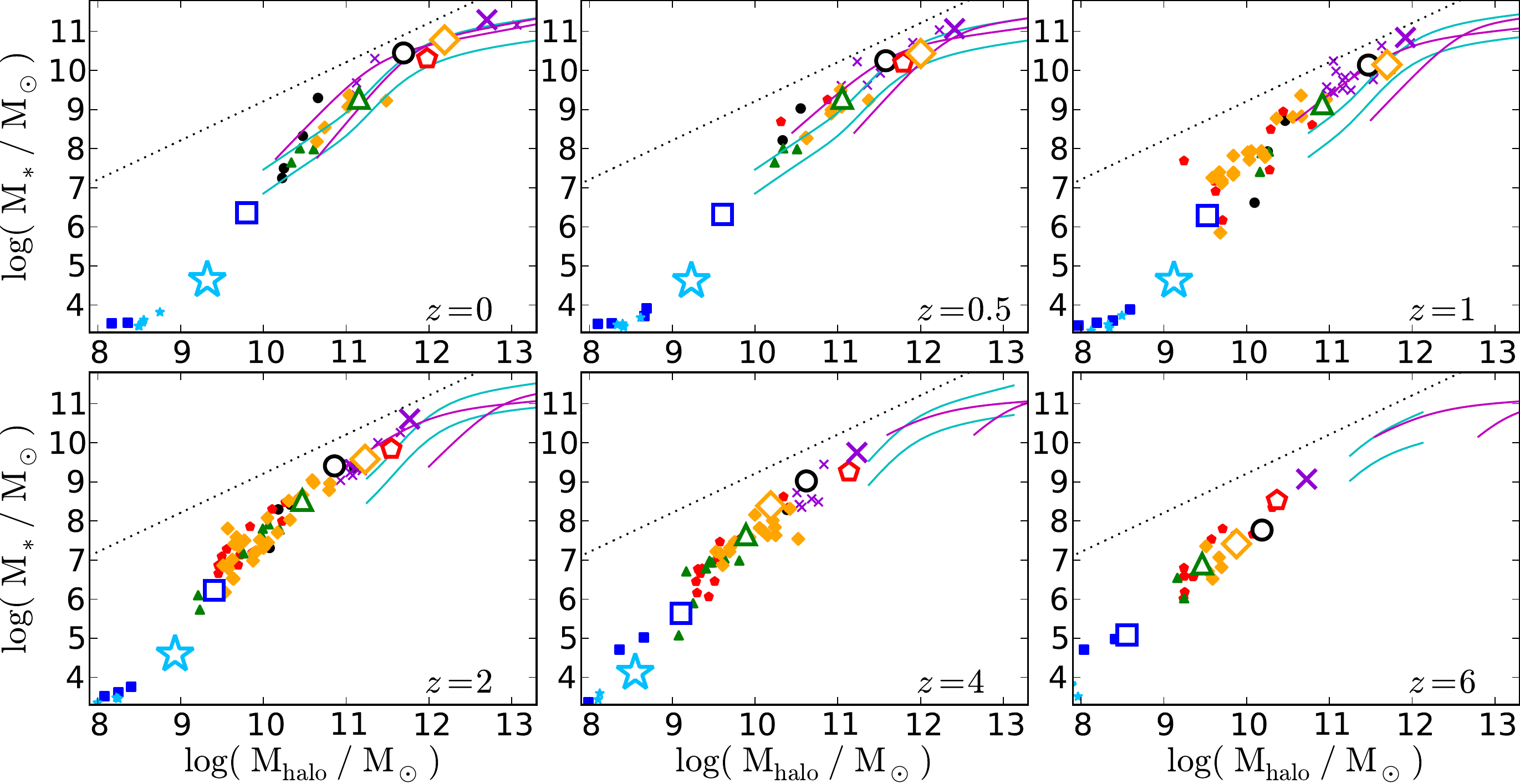}{0.99}
    \vspace{-0.2cm}
    \caption{$M_{\ast}-M_{\rm halo}$ relation as Fig.~\ref{fig:mg.mh.z0} (points follow the legend therein), at different redshifts. 
    Observational constraints are also shown at each redshift (each pair of lines shows the $\pm1\,\sigma$ fit to the observations at that redshift). With no tuned parameters, the simulations predict $M_{\ast}-M_{\rm halo}$ and, by extension, the stellar MF and galaxy clustering, at all $z$. Redshift evolution in $M_{\ast}-M_{\rm halo}$ is weak, with the sense that low-mass dwarfs become higher-$M_{\ast}$, leading to a steeper faint-end galaxy MF, in agreement with constraints from reionization \citep[see][and references therein]{kuhlen:2012.reionization.escape.fractions}.
    \label{fig:mg.mh.z}}
\end{figure*}

The specific halos we re-simulate are chosen to represent a broad mass range and be ``typical'' in most properties (e.g.\ sizes, formation times, and merger histories) relative to other halos of the same $z=0$ mass. The simulations {\bf m09} and {\bf m10} are constructed using the methods from \citet{onorbe:2013.zoom.methods}; they are isolated dwarfs. Simulations {\bf m11}, {\bf m12q}, {\bf m12i}, and {\bf m13} are chosen to match a subset of initial conditions from the AGORA project \citep{kim:2013.AGORA}, which will enable future comparisons with a wide range of different codes. These are chosen to be somewhat quiescent merger histories, but lie well within the typical scatter in such histories at each mass (and each has several major mergers). Simulation {\bf m12v}, for contrast, is chosen to have a relatively violent merger history (several major mergers since $z\sim2$), and is based on the initial conditions studied in \citet{keres:cooling.clumps.from.broken.filaments} and \citet{faucher-giguere:2011.stream.covering.factor}. 

In each case, the resolution is scaled with the simulated mass, so as to achieve the optimal possible force and mass resolution. It is correspondingly possible to resolve much smaller structures in the low-mass galaxies. The critical point is that in all our simulations with mass $<10^{13}\,\msun$, we resolve the Jeans mass/length of gas in the galaxies, corresponding to the size/mass of massive molecular cloud complexes. This is necessary to resolve a genuine multi-phase ISM and for our ISM feedback physics to be meaningful. Fortunately, because most of the mass and star formation in GMCs in both observations \citep{evans:1999.sf.gmc.review,blitz:2004.gmc.mf.universal} and simulated systems (\papertwo) is concentrated in the most massive GMCs, the resolution studies in \paperone-\papertwo\ confirm that resolving small molecular clouds makes little difference. We refer interested readers to \papertwo\ for a detailed discussion of the scales that must be resolved for feedback to operate appropriately, but note here that all our simulations are designed to be approximately comparable to the ``high-resolution'' simulations of isolated galaxies and the ISM in \paperone-\papertwo, within the range of resolution where the results in those studies (star formation rates, wind outflow rates, GMC lifetimes, etc.) were numerically converged (unfortunately, it is not possible to evolve cosmological simulations to $z=0$ with the ``ultra-high'' sub-pc resolution therein). 

In terms of the Jeans mass/length of the galaxies, our resolution is broadly comparable between different simulations. Our worst resolution in units of the Jeans length/mass occurs in the more massive galaxies at late times, when they are relatively gas poor, and so (despite the large total galaxy mass) the Jeans length can become relatively small.\footnote{The approximate Jeans (GMC) mass/length for the $z=0$ disks, assuming Toomre $Q\sim1$, increases from $\sim10^{4}\,\msun$ ($\sim10-30\,$pc) in the $\lesssim10^{10}\,\msun$ halos to $\sim10^{7}\,\msun$ ($\sim100-200\,$pc) in the $\gtrsim10^{12}\,\msun$ halos. If $Q>1$, or if the gas fractions are higher (at higher redshifts), the Jeans masses/lengths are larger as well.} Every galaxy identified in this paper contains at least $\gg 10^{5}$ bound particles. 

We adopt a ``standard'' flat $\Lambda$CDM cosmology with $h\approx0.7$, $\Omega_{{\rm M}}=1-\Omega_{\Lambda}\approx0.27$, and $\Omega_{b}\approx0.046$ for all runs.\footnote{Because of our choice to match some of our ICs to widely-used examples for numerical comparisons, they feature very small cosmological parameter differences. These are percent-level, smaller than the observational uncertainties in the relevant quantities \citep{planck:2013.cosmological.params} and produce negligible effects compared to differences between randomly chosen halos.}

\vspace{-0.5cm}
\section{Baryonic Physics}
\label{sec:sims:physics}

The simulations here use the physical models for star formation and stellar feedback developed and presented in a series of papers studying isolated galaxies \citep{hopkins:stellar.fb.winds,hopkins:stellar.fb.mergers,hopkins:dense.gas.tracers,hopkins:clumpy.disk.evol}, adapted for fully cosmological simulations. We summarize their properties below, but refer to Appendix~\ref{sec:appendix:algorithms} for a more detailed explanation and list of improvements. Readers interested in further details (including resolution studies and a range of tests of the specific numerical methodology) should see \paperone\ \&\ \papertwo.

\vspace{-0.5cm}
\subsection{Cooling}
\label{sec:sims:cooling}

Gas follows an ionized+atomic+molecular cooling curve from $10-10^{10}\,$K, including metallicity-dependent fine-structure and molecular cooling at low temperatures, and high-temperature ($\gtrsim10^{4}\,$K) metal-line cooling followed species-by-species for 11 separately tracked species. At all times, we tabulate the appropriate ionization states and cooling rates from a compilation of {\small CLOUDY} runs, including the effect of the photo-ionizing background, accounting for gas self-shielding. Photo-ionization and photo-electric heating from local sources are accounted for as described below.

\vspace{-0.5cm}
\subsection{Star Formation}
\label{sec:sims:starformation}

Star formation is allowed only in dense, molecular, self-gravitating regions above $n>n_{\rm crit}$ ($n_{\rm crit} = 100\,{\rm cm^{-3}}$ for our primary runs, but we also tested from $\sim10-1000\,{\rm cm^{-3}}$). This threshold is much higher than that adopted in most ``zoom-in'' simulations of galaxy formation (the high value allows us to capture highly clustered star formation). We follow \citet{krumholz:2011.molecular.prescription} to calculate the molecular fraction $f_{\rm H_{2}}$ in dense gas as a function of local column density and metallicity, and allow SF only from molecular gas. We also follow \citet{hopkins:virial.sf} and restrict star formation to gas which is locally self-gravitating, i.e.\ has $\alpha\equiv \delta v^{2}\,\delta r/G\,m_{\rm gas}(<\delta r) < 1$ on the smallest available scale ($\delta r$ being our force softening or smoothing length). This forms stars at a rate $\dot{\rho}_{\ast}=\rho_{\rm mol}/t_{\rm ff}$ (i.e.\ $100\%$ efficiency per free-fall time); so that the galaxy and even kpc-scale star formation efficiency is {\em not} set by hand, but regulated by feedback (typically at much lower values). Because of this, in \paperone, \papertwo, and \citet{hopkins:virial.sf} we show that the galaxy structure and SFR are basically independent of the small-scale SF law, density threshold (provided it is high), and treatment of molecular chemistry.

\vspace{-0.5cm}
\subsection{Stellar Feedback}
\label{sec:sims:feedback}

Once stars form, their feedback effects are included from several sources. Every star particle is treated as a single stellar population, with a known age, metallicity, and mass. Then all feedback quantities (the stellar luminosity, spectral shape, SNe rates, stellar wind mechanical luminosities, metal yields, etc.) are tabulated as a function of time directly from the stellar population models in STARBURST99, assuming a \citet{kroupa:imf} IMF.

(1) {\bf Radiation Pressure:} Gas illuminated by stars feels a momentum flux $\dot{P}_{\rm rad} \approx (1-\exp{(-\tau_{\rm UV/optical})})\,(1+\tau_{\rm IR})\,L_{\rm incident}/c$ along the optical depth gradient, where $1+\tau_{\rm IR} = 1+\Sigma_{\rm gas}\,\kappa_{\rm IR}$ accounts for the absorption of the initial UV/optical flux and multiple scatterings of the re-emitted IR flux if the region between star and gas particle is optically thick in the IR (see Appendix~\ref{sec:appendix:algorithms}). We assume that the opacities scale linearly with gas metallicity.\footnote{There has been some debate in the literature regarding whether or not the full $\tau_{\rm IR}$ ``boost'' applies to the infrared radiation pressure when $\tau_{\rm IR}\gg1$ (see e.g.\ \citealt{krumholz:2012.rad.pressure.rt.instab}, but also \citealt{kuiper:2012.rad.pressure.outflow.vs.rt.method} and \citet{davis:2014.rad.pressure.outflows}, who find much stronger effects in the infrared). We have considered alternatives, discussed in \paperone. However, in the simulations here we never resolve the extremely high densities where $\tau_{\rm IR}\gtrsim1$ (where this distinction is important), and so if anything are {\em under}-estimating the IR radiation pressure, even compared to the most conservative studies.}

(2) {\bf Supernovae:} We tabulate the SNe Type-I and Type-II rates from \citet{mannucci:2006.snIa.rates} and STARBURST99, respectively, as a function of age and metallicity for all star particles and stochastically determine at each timestep if an individual SNe occurs. If so, the appropriate mechanical luminosity and ejecta momentum is injected as thermal energy and radial momentum in the gas within a smoothing length of the star particle, along with the relevant mass and metal yield (for all followed species). When the Sedov-Taylor phase is not fully resolved, we account for the work done by hot gas inside the unresolved cooling radius (converting the appropriate fraction of the SNe energy into momentum). We discuss this in detail in Appendix~\ref{sec:appendix:algorithms}, but emphasize that it is particularly important that SNe momentum not be neglected in massive halos whose mass resolution $\sim10^{4}\,\msun$ is much larger than the ejecta mass of a single SNe. 

(3) {\bf Stellar Winds:} Similarly, stellar winds are assumed to shock locally and so we inject the appropriate tabulated mechanical power $L(t,\,Z)$, wind momentum, mass, and metal yields, as a continuous function of age and metallicity into the gas within a smoothing length of the star particles. The integrated mass fraction recycled in winds (including both fast winds from young stars and slow AGB winds) and SNe is $\sim0.3$.  

(4) {\bf Photo-Ionization and Photo-Electric Heating:} Knowing the ionizing photon flux from each star particle, we ionize each neighboring neutral gas particle (provided there are sufficient photons, given the gas density, metallicity, and prior ionization state), moving outwards until the photon budget is exhausted; this alters the heating and cooling rates appropriately. The UV fluxes are also used to determine photo-electric heating rates following \citet{wolfire:1995.neutral.ism.phases}.

Extensive numerical tests of the feedback models are presented in \papertwo. 

\vspace{-0.5cm}
\section{Simulation Numerical Details}
\label{sec:sims}

All simulations are run using a newly developed version of TreeSPH which we refer to as ``{\small P-SPH}'' \citep{hopkins:lagrangian.pressure.sph}, in the code {\small GIZMO}.\footnote{Details of the {\small GIZMO} code, together with a limited public version, user's guide, movies and test problem examples, are available at\\ \gizmourl}
This adopts the Lagrangian ``pressure-entropy'' formulation of the SPH equations developed in \citet{hopkins:lagrangian.pressure.sph}; this eliminates the major differences between SPH, moving mesh, and grid (adaptive mesh) codes, and resolves the well-known issues with fluid mixing instabilities in previously-used forms of SPH \citep[e.g.][]{agertz:2007.sph.grid.mixing,sijacki:2011.gadget.arepo.hydro.tests}. The gravity solver is a heavily modified version of the {\small GADGET-3} code \citep{springel:gadget}; but {\small GIZMO} also includes substantial improvements in the artificial viscosity, entropy diffusion, adaptive timestepping, smoothing kernel, and gravitational softening algorithm, as compared to the ``previous generation.'' These are all described in detail in Appendix~\ref{sec:appendix:sims}.

We emphasize that our version of SPH has been tested extensively and found to give good agreement with analytic solutions as well as well-tested grid codes on a broad suite of test problems. Many of these are presented in \citet{hopkins:lagrangian.pressure.sph}. This includes Sod shock tubes; Sedov blastwaves; wind tunnel tests (radiative and adiabatic, up to Mach $\sim10^{4}$); linear sound wave propagation; oscillating polytropes; hydrostatic equilibrium ``deformation''/surface tension tests \citep{saitoh:2012.dens.indep.sph}; Kelvin-Helmholtz and Rayleigh-Taylor instabilities; the ``blob test'' \citep{agertz:2007.sph.grid.mixing}; super-sonic and sub-sonic turbulence tests (from Mach $\sim0.1-10^{3}$); Keplerian gas ring, disk shear, and shearing shock tests \citep{cullen:2010.inviscid.sph}; the Evrard test; the Gresho-Chan vortex; spherical collapse tests; and non-linear galaxy formation tests. For additional tests showing the improvements relative to previous-generation SPH, see \citet{hu:2014.psph.galaxy.tests}. 
Since it is critical for the problems addressed here that a code be able to handle high dynamic range situations, the numerical method and parameters such as SPH ``neighbor number'' were not modified for these tests individually, but are similar to what we use in our production runs in this paper.

In Appendix~\ref{sec:appendix:sims}, we note that we have explicitly tested many of the purely numerical elements of the gravity and hydrodynamic solvers in the simulations shown here: for example, whether to use adaptive or fixed gravitational softenings, the choice of SPH smoothing kernel, and the timestepping algorithm. However we do not discuss these in the main text because they produce extremely small ($\lesssim 10\%$-level) differences in the quantities plotted in this paper.

\vspace{-0.5cm}
\section{Results}
\label{sec:results}

\subsection{Galaxy Masses as a Function of Redshift}
\label{sec:results:masses}

Fig.~\ref{fig:mg.mh.z0} plots the $z=0$ stellar mass-halo mass relation for our main set of simulations from Table~\ref{tbl:sims} (highest-resolution, with all physics enabled). Note that although each high-resolution region at $z=0$ contains one ``primary'' halo (the focus of that region), there are several smaller-mass, independent halos also in that region. We therefore identify and plot all such halos.\footnote{We use the {\small HOP} halo finder \citep{eisenstein:1998.hop.halo.finder} to automatically identify halos (which combines an iterative overdensity identification with a saddle density threshold criterion to merge subhalos and overlapping halos). Halo masses are defined as the mass within a spherical aperture about the density maximum with mean density $>200$ times the critical density at each redshift (this is chosen to be similar to the choice used in abundance-matching models, which define the observations to which we compare). Stellar mass plotted is the total stellar mass within $20\,$kpc of the center of the central galaxy in the halo (we do not include satellite galaxy masses). However, we have compared with the results of a basic friends-of-finds routine or simple by-eye identification, and find that for the results here (focused on simple, integral halo quantities, and ignoring subhalos), this makes no significant difference. Likewise defining the mass within $\sim0.1\,R_{\rm vir}$ instead of $20\,$kpc makes no significant difference.}
We exclude those halos that are outside the high-resolution region (more than $1\%$ mass-contaminated by low-resolution particles; although varying this between $0.5-10\%$ makes little difference to our comparisons here) or insufficiently resolved ($<0.01$ times the primary halo mass, or with $<10^{5}$ dark matter particles). We also exclude subhalos/satellite galaxies.

The known sources of stellar feedback we include, with {\em no} adjustment, automatically reproduce a relation between galaxy stellar and halo mass consistent with the observations\footnote{\label{foot:moster.behroozi.halo.defns}Note that \citet{behroozi:2012.abundance.matching.sfhs} and \citet{moster:2013.abundance.matching.sfhs} use definitions of halo mass which differ slightly (by $\approx10\%$). For our purposes, this produces negligible differences in our comparison.} from $M_{\rm halo}\sim 10^{7}-10^{13}\,\msun$. Specifically, the distribution of points for all $M_{\rm halo}\le 10^{12}\,\msun$ is statistically consistent (in a $\chi^{2}$ sense) with having been drawn from the \citet{moster:2013.abundance.matching.sfhs} curve; if we allow for the observed scatter ($\sim0.15\,$dex at $\sim 10^{12}\,\msun$, the width between the plotted lines, increasing to $\sim0.3\,$dex at the lowest observed masses) then all our primary galaxies lie within the $2\,\sigma$ scatter.\footnote{We can also fit the points here to the same power-law functional form used empirically: if we do so, the best-fit slope and normalization are both within $0.5\,\sigma$ of the fit to observations in \citet{moster:2013.abundance.matching.sfhs} (the error bar is dominated by the small-number statistics in our halo sampling). The simulations with $M_{\rm halo}\ll 10^{10}\,\msun$ are statistically inconsistent with the extrapolation of the flatter slope from \citet{behroozi:2012.abundance.matching.sfhs}, but this is entirely below the region actually observed, where \citet{behroozi:2012.abundance.matching.sfhs} and \citet{moster:2013.abundance.matching.sfhs} agree well, and there the simulations do not significantly ``prefer'' either fit.}

Despite the fact that this relation implies a non-uniform (and even non-monotonic) efficiency of star formation as a function of galaxy mass, we do {\em not} need to invoke different physics or distinct parameters at different masses. This is particularly impressive at low masses, where the integrated stellar mass must be suppressed by factors of $\sim1000$ relative to the Universal baryon fraction. Unfortunately, at high masses ($>10^{13}\,\msun$), the large Lagrangian regions (hence large number of required particles) limit the resolution we can achieve; we have experimented with some low-resolution test runs which appear to produce overly massive galaxies, but higher-resolution studies are required to determine if that owes to a need for additional physics or simply poor numerical resolution. 

Interestingly, the scatter in $M_{\ast}$ at fixed $M_{\rm halo}$ may decrease weakly with mass, from $\sim0.5$\,dex in dwarf galaxies ($M_{\rm halo}\lesssim 10^{10}\,\msun$) to $\sim0.1-0.2$ dex in massive ($\sim L_{\ast})$ galaxies. But given the limited number of halos we study here, further investigation allowing more diverse merger/growth histories is needed.

Fig.~\ref{fig:mg.mh.z} shows the $M_{\ast}-M_{\rm halo}$ relation at various redshifts. At each $z$, we compare with observationally constrained estimates of the $M_{\ast}-M_{\rm halo}$ relation. Implicitly, if they agree in $M_{\ast}(M_{\rm halo})$, our models are consistent with the observed stellar MF (given, of course, the limited statistics by which we are ``sampling'' the MF). 
At high redshifts, the halos we simulate are of course lower-mass, so eventually we have no high-mass galaxies; this limits the extent to which our results can be compared to observations above $z\sim2$. 

\vspace{-0.5cm}
\subsection{Other (Basic) Galaxy Properties}

We wish to focus here on galaxy masses and star formation histories. Companion papers (in preparation) will examine the galaxy morphologies and other observables in more detail. It is important, when studying those properties, to construct a meaningful comparison (e.g.\ using the same methods and wavelengths observed), and this is non-trivial. Moreover, it is by no means obvious that these properties are as robust to numerical details as the galaxy stellar masses (discussed further below), and it is completely outside the scope of this paper to fairly explore those dependencies.

That said, we can briefly note the basic properties of the specific simulations in Table~\ref{tbl:sims} at $z=0$, with the caveat that these {\em may not be robust} to changes in either the initial conditions (the particular halo simulated) or our numerical methods. Morphologically, at $z=0$, run {\bf m09} resembles an ultrafaint dwarf;  {\bf m10} a thick, but rotating dwarf irregular; and {\bf m11} a more ``fluffy'' dwarf spheroidal. Runs {\bf m12v}, {\bf m12q}, {\bf m12i} produce bulge+disk systems, with {\bf m12v} showing a prominent bulge at all times $z\lesssim2$; {\bf m12q} is more disk-dominated until a late major merger at $z< 0.5$ destroys the disk; and run {\bf m12i} produces a stellar disk with little bulge. Run {\bf m13} is totally bulge-dominated. Each galaxy has an approximately flat rotation curve outside of the central couple kpc; those with $M_{\ast}<10^{10}$ slowly rise with radius to $V_{\rm max}$, and the more massive systems are flat to within the central $\sim$\,kpc, except for {\bf m12v} (where the compact bulge leads to a central-kpc spike at $\sim 250\,{\rm km\,s^{-1}}$). The galaxy sizes, measured as the half-stellar mass effective radii, are $(0.3,\,0.52,\,3.5,\,2.8,\,3.2,\,4.2,\,4.8)\,{\rm kpc}$ for ({\bf m09, m10, m11, m12v, m12q, m12i, m13}), consistent with the observed stellar size-mass relation \citep{shen:size.mass,wolf:2010.disperson.gal.masses}.

\vspace{-0.5cm}
\subsection{(Lack of) Dependence on Numerical Methods}
\label{sec:results:numerics}

\begin{figure}
    \centering
    \plotonesize{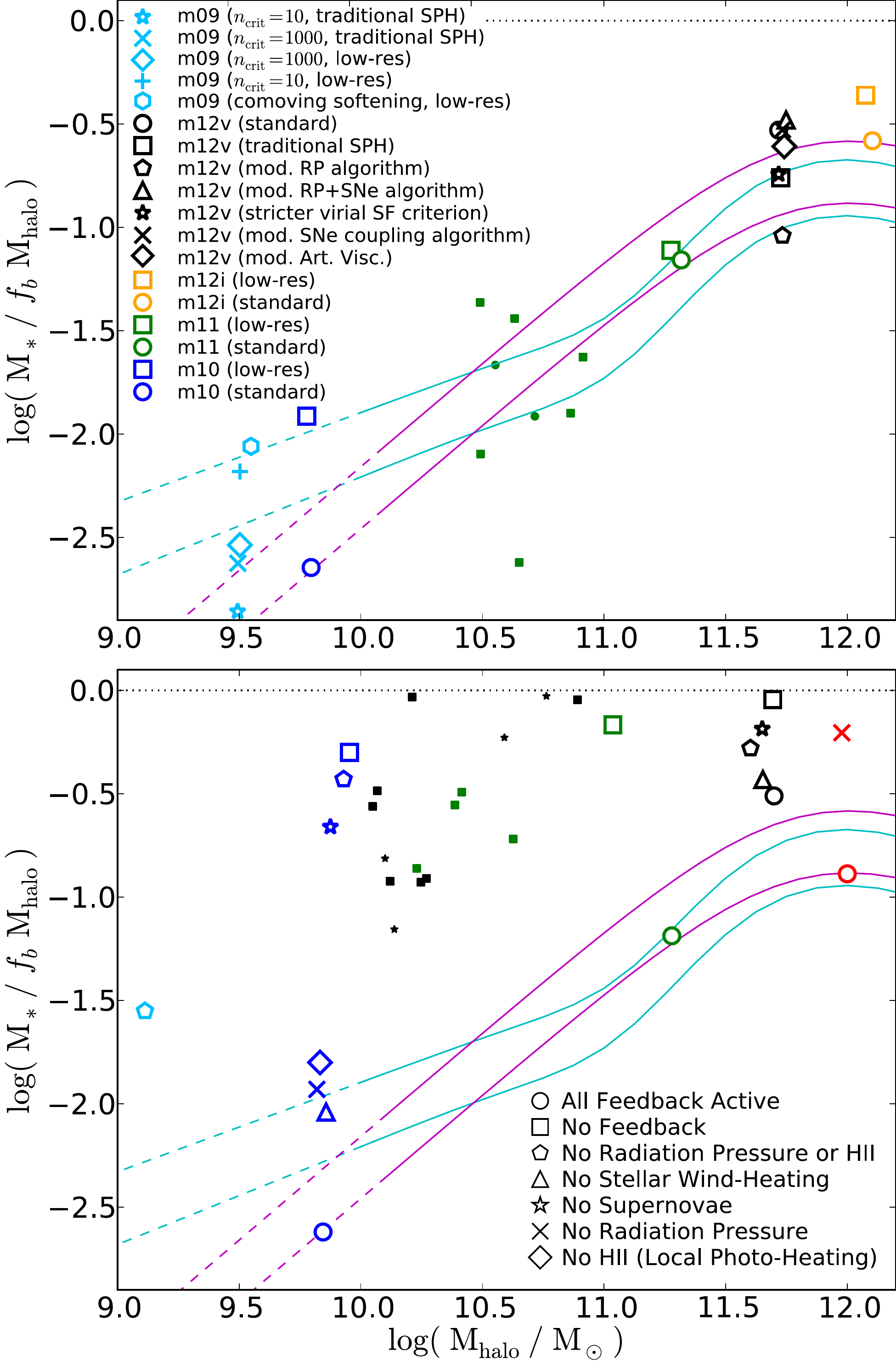}{1.01}
    \vspace{-0.5cm}
    \caption{$M_{\ast}-M_{\rm halo}$ relation at $z=0$, as Fig.~\ref{fig:mg.mh.z0}. 
    {\em Top:} Simulations with different numerical parameters: we show the effects of varied resolution, artificial viscosity, and the algorithmic implementation of feedback. We also compare a completely different version of SPH (with a different set of hydrodynamic equations), which is known to differ significantly in certain idealized hydrodynamic test problems. These have little effect on our predictions.
    {\em Bottom:} Effect of physical variation in stellar feedback properties. We compare runs with no stellar feedback, with no supernovae (but stellar winds, radiation pressure, and photo-ionization heating included), or with no radiative feedback (radiation pressure and local HII-heating). ``No feedback'' runs generally predict $M_{\ast}\sim f_{b}\,M_{\rm halo}$, in severe conflict with the observations.$^{\ref{foot:nofb.lowmhalo}}$ Removing radiative {\em or} SNe feedback also produce order-of-magnitude too-large stellar masses. The non-linear {\em combination} of feedback mechanisms (not any one in isolation) is critical to drive winds and regulate galaxy masses.
    \label{fig:mg.mh.fb}}
\end{figure}

In Fig.~\ref{fig:mg.mh.fb} we investigate how the $M_{\ast}-M_{\rm halo}$ relation depends on numerical parameters and feedback. First we repeat Fig.~\ref{fig:mg.mh.z0} for simulations with different {\em purely numerical} parameters. These can and do, indeed, have significant quantitative effects -- they can easily shift the predicted stellar masses by factors $\sim2-3$. However, we stress that they do not {\em qualitatively} change our conclusions. 

Modest changes in resolution (our ``low-resolution'' runs correspond to one power of two step in spatial resolution, and a corresponding factor of $2^{3}=8$ change in mass resolution) lead to significant, but not order-of-magnitude, changes in $M_{\ast}$: generally we obtain larger $M_{\ast}$ by factors of $\sim1.5$ at high masses ($M_{\rm halo}\gtrsim 10^{11}\,\msun$) and $\sim2-3$ at the lowest masses ($M_{\rm halo}<10^{10}\,\msun$) at lower resolution, owing to a combination of (a) artificially enhanced mixing and thus cooling of diffuse gas, since ISM phases are less well-resolved, and (b) the fact that the coupling of feedback energy and momentum is necessarily spread over larger mass elements. If we downgrade our resolution more substantially -- by a factor of $\sim100$ in mass, or $>10$ in spatial scale (i.e.\ using the $>100\,$pc spatial resolution which is typical of most previous cosmological simulations), the results diverge more substantially: galaxy masses at $z\sim0$ are a factor of $\sim3-5$ higher at high masses and $\sim10$ higher at low masses. This makes sense, because at that resolution, we simply cannot meaningfully resolve even the most massive structures in the ISM.\footnote{We have run a couple tests with $30$ times higher particle numbers than our production-quality runs (for {\bf m12i} and {\bf m10}), to $z=2$, and found that the stellar masses at this time and earlier vary by $\sim 10-50\%$ from those quoted here. However, this appears to be primarily stochastic, rather than systematic, so we suspect the masses will not change much further at still higher resolution.}

Some of our numerical tests are not plotted here because their effects are not significant. We have, for example, re-run several simulations with twice and five times larger dark matter softening lengths (same baryonic softening); using or de-activating adaptive gravitational softenings (which ensure there are always $\sim100$ neighbor particles in the softening kernel); varying the number of SPH ``neighbors'' in the hydrodynamic kernel and number of SPH particles to which energy and momentum are coupled; using a single timestep or Strang-split integration scheme in the code; varying the Courant factor of the hydrodynamic solver; changing the order of operator-splitting for the cooling and feedback steps; or forcing equal vs.\ allowing separate gravitational softenings for baryons and dark matter. These produce very small ($<10\%$) differences. We also varied the sizes of the high-resolution ``zoom-in'' Lagrangian regions of the halos; the results here are insensitive to the region size if we choose sizes $\gtrsim2\,R_{\rm vir}$ (at the redshift of interest), but the cooling of halo gas and star formation are artificially suppressed if the high-resolution region is much smaller. 

Changing the small-scale star formation prescription in the simulations has very little effect on our predictions. This is expected based on all of our previous studies using isolated (non-cosmological) simulations (for explicit examples where we vary the density threshold and instantaneous ``efficiency'' of star formation in dense gas by factors of $>1000$, as well as the density, temperature, chemical, and virial-state dependence of star formation, see \paperone, \papertwo, and \citealt{hopkins:virial.sf}). Globally, star formation is {\em feedback-regulated}: a certain number of young stars are required to balance gravitational collapse/dissipation, independent of how those stars form. So long as cooling can proceed, they will form (what will change, via this self-regulation, if we change e.g.\ the density threshold above which stars form, is the amount of gas which ``builds up'' above that threshold; see \citealt{hopkins:dense.gas.tracers}). In Fig.~\ref{fig:mg.mh.fb} we show examples where we vary the density threshold for star formation by factors of $\sim100$ (producing $<0.2$\,dex random/non-systematic changes in stellar mass), or impose a much stricter local virial criterion for star formation (local virial parameter $<0.5$ instead of $1$; producing a $\sim20\%$ difference in stellar mass); we have also experimented with removing the virial parameter entirely or dividing the instantaneous efficiency of star formation in dense gas by a factor of $100$ (both produce $<10\%$ changes). 

We have also investigated different purely algorithmic methods for coupling the same feedback physics. Subtle differences in the algorithmic implementation of feedback have little {\em systematic} effect on the stellar mass, provided the same mechanisms are included; however they can only be compared statistically, since the stochastic nature of feedback means that even very subtle changes can produce significant differences in the exact time history of bursts, for example. Of what we have considered, the most important parameter is how we implement the momentum gained during the Sedov-Taylor phase of SNe remnant expansion when the cooling radius is unresolved (see Appendix~\ref{sec:appendix:algorithms}). For example, one of the ``mod.\ SNe coupling algorithm'' examples changes the particle weights (using a standard SPH kernel weight -- effectively mass-weighted in the smoothing kernel -- instead of a volumetric weighting) used to determine the coupling of SNe energy and momentum in the kernel. This can have dramatic effects on test problems: for a SNe in an infinitely thin, adiabatic disk with a low-density exterior, a mass-weighting couples all the momentum in the disk plane, instead of the vertical direction (the correct solution). Nevertheless we see this has relatively weak ($\sim20\%$) effects on the stellar mass and star formation history (in part because, in the average over many SNe over large volumes, all that matters is the total feedback input); however, it can significantly effect the morphological structure of e.g.\ the dense gas in a thin disk. We have also experimented with different functional forms for the ratio of the SNe energy and momentum coupling (producing small effects). The ``mod.\ RP algorithm'' choice discretizes our radiation pressure term (which is usually a continuous force) into intentionally very large ($>500\,{\rm km\,s^{-1}}$) ``kicks'' (this keeps the same total momentum flux, but makes each such particle ``kicked'' unbound) -- unsurprisingly this suppresses star formation further, but only by a factor of $\sim 3$. In the ``mod.\ RP+SNe'' choice we discretize the radiation pressure into smaller kicks ($=5\,{\rm km\,s^{-1}}$) and see this has little effect (as expected). In all cases the results lie within the (rather large) range allowed by observations. 


In a companion paper, \citet{keres:2013.fire.cosmo.vs.numerics} consider the detailed effects of substantial changes to each aspect of our numerical method described in Appendix~\ref{sec:appendix:sims}. Here, we simply show a few basic comparisons. Considerable attention has recently been paid to differences between the results of grid codes and older SPH methods (such as that in \citealt{springel:entropy}) for certain problems (especially sub-sonic fluid mixing instabilities; see \citealt{agertz:2007.sph.grid.mixing,kitsionas:2009.grid.sph.compare.turbulence,bauer:2011.sph.vs.arepo.shocks,vogelsberger:2011.arepo.vs.gadget.cosmo,sijacki:2011.gadget.arepo.hydro.tests,keres:2011.arepo.gadget.disk.angmom}). The numerical method used for our standard simulations has been specifically shown to resolve most of these discrepancies (giving results quite similar to grid codes in test problems); this is verified in \citet{hopkins:lagrangian.pressure.sph} for standard test problems and \citet{keres:2013.fire.cosmo.vs.numerics} for cosmological simulations. However we have re-run some of our simulations using the \citet{springel:entropy} formulation of SPH (described in Appendix~\ref{sec:appendix:sims}), which shows the most pronounced forms of these discrepancies. Despite the known differences between such methods for certain test problems, we find in Fig.~\ref{fig:mg.mh.fb} ({\em top} panel) that it makes little difference for the predicted galaxy masses. The older SPH method gives slightly lower $M_{\ast}(M_{\rm halo})$ (by about $\approx0.15\,$dex), primarily because cooling of diffuse ``hot halo'' gas is suppressed by less-efficient mixing. But for this specific question, the effect is quite small compared to the effects of including the appropriate stellar feedback physics. We also show an experiment where adopt an entirely distinct artificial viscosity prescription (see Appendix~\ref{sec:appendix:sims} for details), which produces negligible differences.

It is important to stress that our conclusion here -- that our results depend only weakly on the numerical details -- applies to the galaxy stellar masses and other lowest-order, integrated quantities. In future work, we will study other properties of the simulations, such as the galaxy morphologies, which can (and do) depend on some parameters much more sensitively. For example, the modifications to the SNe coupling algorithm described above, which produce very little systematic change in our predicted stellar masses and star formation histories, produce surprisingly large changes to the angular momentum content and thickness of disks in the more massive galaxies. 

Finally, we show these results in part to stress, emphatically, that while there are {\em always} numerical choices in any code, there has been no ``tuning'' of these parameters for our study here. Certainly none of these has been ``fit'' or ``adjusted'' to match any observations, and all the choices above are held constant across our standard set of simulations, using values calibrated from simple test problems \citep[e.g.][]{hopkins:lagrangian.pressure.sph}.

\begin{figure}
    \centering
    \plotonesize{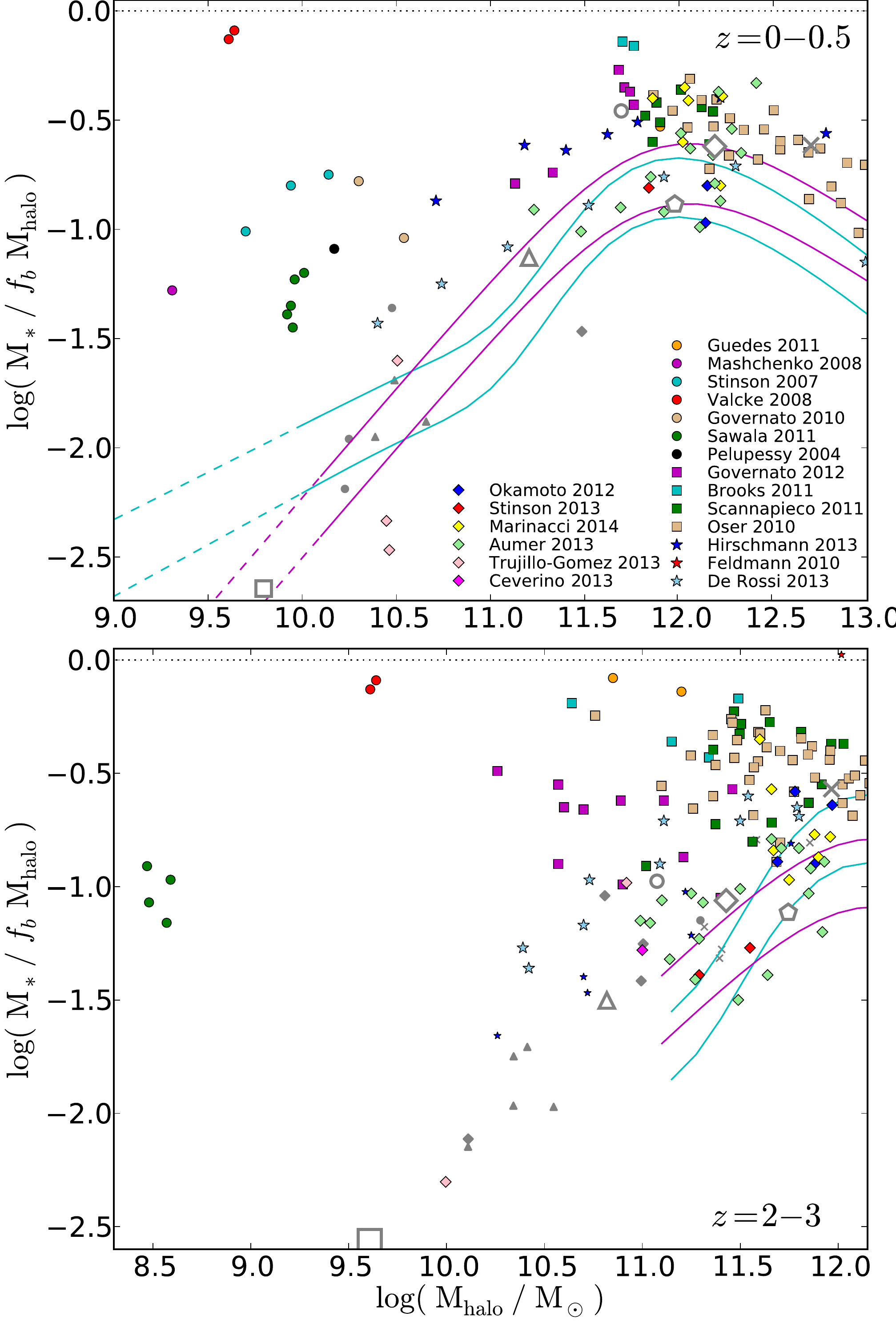}{1.01}
    \vspace{-0.5cm}
    \caption{Comparison of the $M_{\ast}(M_{\rm halo})$ relation (as Fig.~\ref{fig:mg.mh.z0}) predicted by other published simulations in the literature using sub-grid stellar feedback models. We compile these results (colored points), where available, at low-$z$ ($z=0-0.5$; {\em top}) and high-$z$ ($z=2-3$; {\em bottom}). We compare against the simulations presented here (gray points) with explicit feedback, and observational constraints (lines). Even sub-grid models which are ``successful'' near $\sim L_{\ast}$ at $z\sim0$ over-predict $M_{\ast}(M_{\rm halo})$ by an order-of-magnitude relative to our explicit feedback simulations and observations at both low masses ($M_{\rm halo}\lesssim10^{10}\,\msun$) and/or high redshifts ($z\gtrsim2$). The exceptions appear to be the newest generation of sub-grid models which have been explicitly adjusted to mimic the effects of radiative feedback as well as SNe, seen in our explicit feedback models: this includes \citet{stinson:2013.new.early.stellar.fb.models,aumer:2013.new.cosmo.zooms.rad.pressure.fb,ceverino:2013.rad.fb,trujillo-gomez:2013.rad.fb.dwarfs}.
    \label{fig:mg.mh.altsims}}
\end{figure}

\vspace{-0.5cm}
\subsection{(Strong) Dependence on Feedback}
\label{sec:results:feedback}

The lower panel in Fig.~\ref{fig:mg.mh.fb} shows the effect of varying the {\em physics} of feedback: now, we see dramatic differences in $M_{\ast}(M_{\rm halo})$. Removing all feedback (every mechanism listed in \S~\ref{sec:sims:feedback}), gas cools and collapses on a dynamical time $t_{\rm dyn}$ within the disk, forming stars at a rate $\dot{M}_{\ast} \sim M_{\rm gas}/t_{\rm dyn} \sim \dot{M}_{\rm gas}$ where $\dot{M}_{\rm gas}$ is the inflow rate from the halo. Most of the baryons are turned into stars.\footnote{\label{foot:nofb.lowmhalo}Even with no feedback, at very low masses $M_{\rm halo}\lesssim 10^{9}\,\msun$, some suppression of SF occurs after reionization because we still include a photo-ionizing background. However the predicted stellar mass is still larger than observed by at least an order of magnitude.}

If we turn off SNe feedback, but retain all other forms of feedback, the results are nearly as bad: again, $M_{\ast}$ is severely overpredicted in both dwarfs and MW-mass systems. In the $\lesssim 10^{10}$ halos, with no SNe, other forms of feedback may still suppress SF significantly (so $M_{\ast}\ll f_{b}\,M_{\rm halo}$), but the masses are still much too large relative to those observed by factors of $\sim 100$. We also note that, as many previous studies have pointed out \citep{murray:momentum.winds,mckee:2007.sf.theory.review,shetty:2008.sf.feedback.model,cafg:sf.fb.reg.kslaw}, it is ultimately the {\em momentum} injected by SNe, not just the thermal energy, which regulates star formation. So as expected, if we artificially turn off the SNe momentum (coupling only thermal energy, as is common in many cosmological simulations), then in our simulations of massive ($>10^{12}\,\msun$) halos, this is nearly as bad as removing SNe entirely. In the lowest-mass dwarfs, the discrepancy is not so severe (factor $\lesssim2$ changes in the SFH), because the mass resolution ($\sim 100\,\msun$) is such that the early expansion phases of SNe remnants (in which the thermal energy begins to be converted into momentum) are well-resolved.

If we remove radiative feedback entirely (both radiation pressure and local photo-ionization and photo-electric heating, as described in \S~\ref{sec:sims:feedback}), but retain SNe (and stellar winds), we see a nearly identical failure (to the no-SNe case) in both dwarfs and massive galaxies: while $M_{\ast} < f_{b}\,M_{\rm halo}$, far too many stars form. As we showed in \papertwo, these mechanisms are critical to disrupt the dense regions of GMCs in which young stars are born, {\em before} SNe explode, and thus allowing the SNe to heat larger, lower-density volumes of gas (which can both avoid over-cooling and feel the collective effects of many SNe rather than just one), and therefore actually generate significant galactic outflows. The same result is found (on smaller scales) in much higher-resolution simulations of either single star clusters or the first stars, which directly treat the radiation-hydrodynamics with each single star as a source \citep[e.g.][]{offner:2009.rhd.lowmass.stars,krumholz:2011.rhd.starcluster.sim,tasker:2011.photoion.heating.gmc.evol,wise:2012.rad.pressure.effects}. 

Interestingly, in the dwarfs, if we turn off {\em only} radiation pressure, or {\em only} photo-ionization heating, the effect is much less severe: the predicted stellar mass is still significantly larger, but it is $>100$ times larger when both are removed. Radiation pressure can, to some extent, ``make up for'' the loss of photo-heating, and vice versa. This should actually not be surprising: the most massive GMCs in dwarf galaxies have local characteristic velocities $< 10\,{\rm km\,s^{-1}}$, thus either HII heating or UV radiation pressure alone can disrupt them (though we expect, under these conditions, HII heating should dominate), and this is completely consistent with both observations of star-forming regions \citep[e.g.][]{lopez:2010.stellar.fb.30.dor} and numerical radiation-hydrodynamic simulations of low-density, low-velocity clouds \citep{harper-clark:2011.gmc.sims,sales:2013.phototion.fb.strong}. And indeed this tradeoff between photo-heating and radiation pressure in small clouds is exactly what we saw in our ultra-high resolution simulations of isolated dwarfs of the same mass (discussed extensively in \papertwo; see Figs.~7, 9, and 14-19 therein). 

In the massive systems, on the other hand, the radiation pressure term becomes more important than the HII heating. We see this in tests with both {\bf m12q} and {\bf m12v}. Even when the difference in stellar mass is not large (e.g.\ the {\bf m12v} case), the lack of radiation pressure feedback is particularly evident in the dense, early-forming center of the galaxy, where in the runs without radiation pressure feedback an enormous central density ``spike'' appears, leading to a very large circular velocity of $\sim 400\,{\rm km\,s^{-1}}$ in the central regions of these systems. At these densities, HII photo-heating is dynamically insignificant.

If we disable stellar wind feedback (specifically, retaining stellar winds as a source of mass and metals, but associating no energy or momentum with that mass), and retain all other feedback, we see relatively weak effects. This is not surprising: their momentum flux is comparable to but not larger than other sources, and their energetics are much less than SNe. But they are obviously an extremely important source of mass and metals in the ISM.

\vspace{-0.5cm}
\subsection{Comparison to Previous Work}
\label{sec:results:previous}

In Fig.~\ref{fig:mg.mh.altsims}, we compare our results (grey points) at low and high redshifts, to those from previous simulations spanning a wide range of galaxy properties and numerical methods \citep{pelupessy:2004.dwarf.gal.sims.burst.sfhs,
stinson:2007.dwarf.gal.sims.burst.sfh,stinson:2013.new.early.stellar.fb.models,
mashchenko:2008.dwarf.sne.fb.cusps,
valcke:2008.dwarf.gal.cosmo.sims,
governato:2010.dwarf.gal.form,governato:2012.dwarf.form.sims,
oser:2010.twophase.galform.sims,
feldmann:bgg.size.evol.in.hydro.sims,
brooks:2011.disk.scaling.law.sims,
guedes:2011.cosmo.disk.sim.merger.survival,
sawala:2011.dwarf.gal.sims.cores,
scannapieco:2011.mw.mass.gal.form,
de-rossi:2013.dwarf.cosmo.sims,
okamoto:2013.pseudobulge.cosmo.sim,
kannan:2013.early.fb.gives.good.highz.mgal.mhalo}. {\em All} of these simulations include some form of sub-grid model designed to mimic the ultimate effects of stellar feedback, although the prescriptions adopted differ substantially between each. Most of these models are specifically tuned to reproduce reasonable scaling for MW-mass systems at $z\sim0$. However, two discrepancies are immediately evident. First, nearly all the previous models predict much larger stellar masses in dwarf galaxies with $M_{\rm halo}\lesssim10^{11.5}\,\msun$, compared to either our simulations or the observational constraints. Second, even simulations which produce excellent agreement with the observations at $z=0$ tend to predict far too much star formation at high redshift (take e.g.\ the simulation in \citealt{guedes:2011.cosmo.disk.sim.merger.survival}, which produces a MW-like system with many properties consistent with observations at $z=0$, but has turned nearly all its baryons into stars at $z\gtrsim2$). 

These are similar to the discrepancies that appear when we re-run our simulations excluding radiative feedback. And indeed, nearly all of the models from the literature in Fig.~\ref{fig:mg.mh.altsims}, even given various freely adjustable parameters, are designed and motivated only to reproduce the effects of supernova feedback, which we have shown is insufficient to explain the observations. 

In fact, the only sub-grid models, to our knowledge, which currently do not produce such discrepancies (and agree broadly with our simulations both at low masses and high redshifts) are the recent generation of models in \citet{stinson:2013.new.early.stellar.fb.models,aumer:2013.new.cosmo.zooms.rad.pressure.fb,ceverino:2013.rad.fb,trujillo-gomez:2013.rad.fb.dwarfs} (for some additional results from these see \citealt{kannan:2013.early.fb.gives.good.highz.mgal.mhalo}). These new models (all of which have been developed recently) are {\em specifically} designed/tuned to mimic the effects of radiative feedback (albeit indirectly), and to reproduce via simple sub-resolution prescriptions (including turning off cooling) some of the most important effects of radiation pressure and photo-heating which were studied in our previous work \citep{hopkins:fb.ism.prop}.\footnote{There have also been interesting results from the re-tuned wind model of \citet{oppenheimer:outflow.enrichment} used more recently in slightly different forms in \citet{torrey:2013.arepo.cosmo.sim.vs.obs}, \citet{marinacci:2014.disk.form.arepo.subgrid}, and \citet{hirschmann:2013.mgal.mhalo.subgrid}. However, in this model, the wind outflow rates are set {\em explicitly} by-hand (and in fact the most recent scalings used were adjusted based on comparison to the sub-grid models including radiative feedback), and then tuned to reproduce the observed mass function. So this is essentially what we attempt to {\em predict} here.} Whether this is unique or not remains to be tested; the phase structure and other properties of outflows and the CGM in such models can be very different from those predicted here, even for the same mass-loading efficiencies (discussed further below). It will be particularly interesting to see whether other recently-developed sub-grid models such as that in \citet{agertz:2013.new.stellar.fb.model}, also incorporating the effects of radiative feedback but via very different prescriptions, will also agree well with observations at both low and high redshifts. In any case, these comparisons -- and the results from this new generation of sub-grid models -- highlight that some accounting for non-SNe feedback is critical.


\begin{figure}
    \centering
    \plotonesize{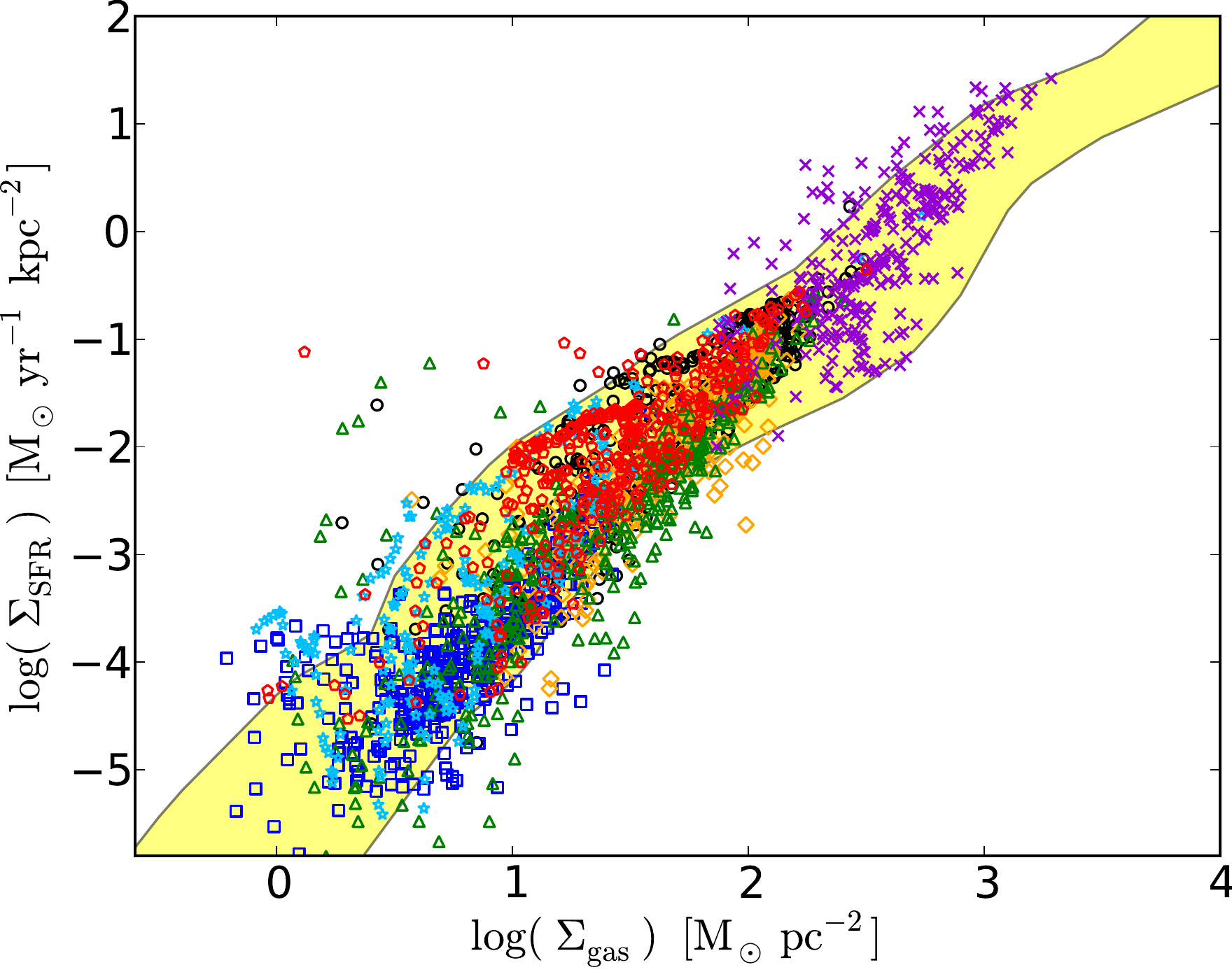}{1.01}
    \vspace{-0.5cm}
    \caption{Kennicutt-Schmidt law, observed \citep[][yellow shaded range]{kennicutt98,bigiel:2008.mol.kennicutt.on.sub.kpc.scales,genzel:2010.ks.law,daddi:2010.ks.law.highz} and simulated (points as Fig.~\ref{fig:mg.mh.z0}). We emphasize that this is a prediction: the instantaneous SF efficiency per dynamical time in dense gas is $100\%$ in the simulations, but the emergent KS-law, as a consequence of feedback, has an efficiency a factor $\sim50$ lower. As shown in \paperone\ and \citet{hopkins:virial.sf}, this is insensitive, with resolved feedback models, to the small-scale star formation law, and entirely determined by stellar feedback. ``No feedback'' models lie a factor $\sim50$ above the observations; ``no radiation'' and ``no SNe'' models (Fig.~\ref{fig:mg.mh.fb}) lie a factor $\sim10$ above observations.
    \label{fig:ks}}
\end{figure}

\begin{figure}
    \centering
    \plotonesize{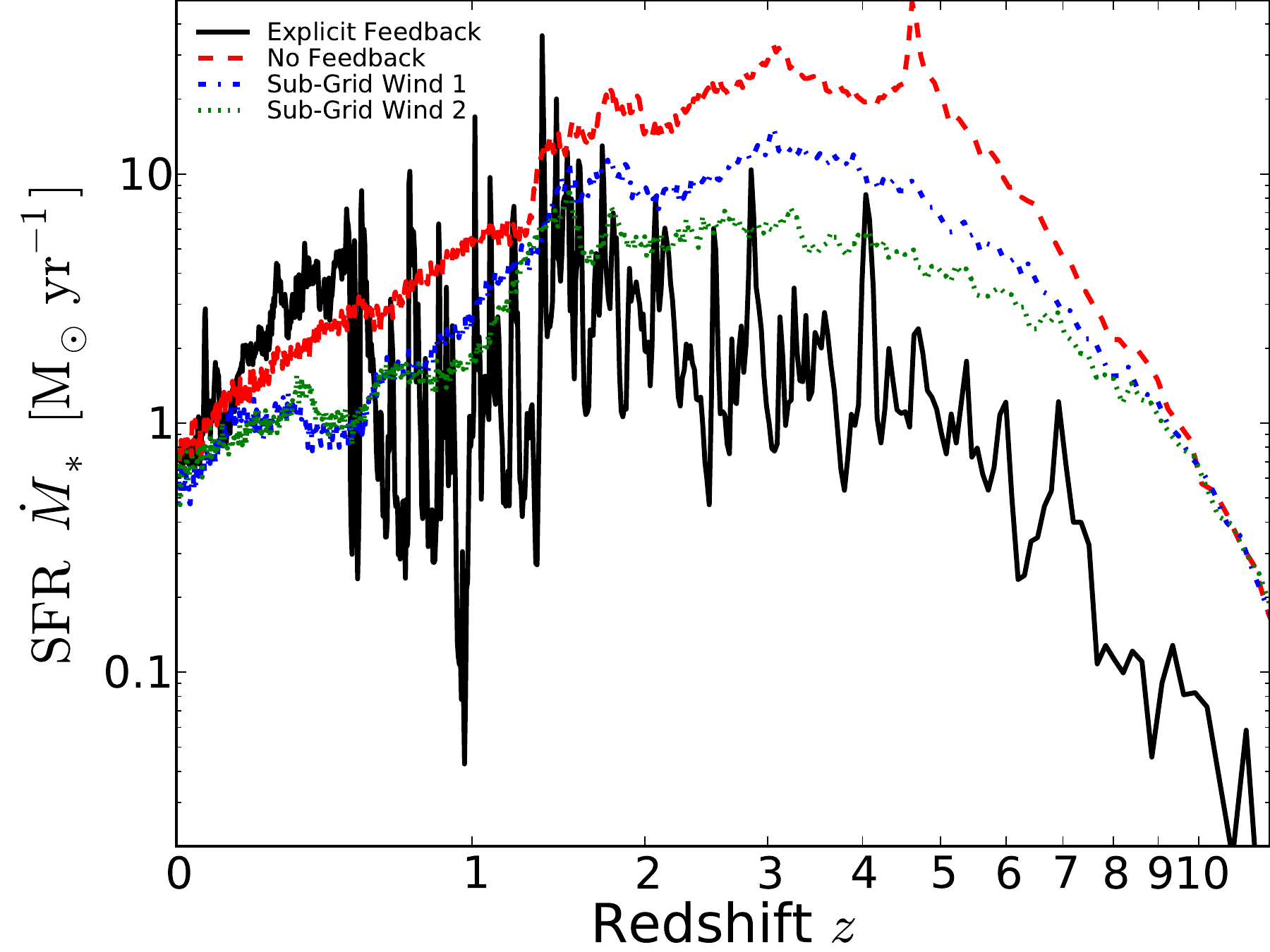}{1.01}
    \caption{Example star formation history (SFH) for the {\bf m12v} simulation, in our standard (explicit feedback) model compared to different sub-grid feedback treatments. We show the formation history of all stars in the simulated box at $z=0$ (smoothed in $10^{7}$\,yr bins); this is not qualitatively different from the SFR versus time of the largest ``main'' galaxy in the box at each time. ``No feedback'' models force the galaxy to lie on the KS-law (SF is ``slow'') but do not expel gas; gas piles up until the SFR balances the halo accretion rate, with a broad peak from $z\sim2-6$. ``Sub-grid wind'' models ``kick'' gas at a rate proportional to the SFR; we show examples with different efficiencies and implementations. By design, model ``2'' produces a nearly identical $z=0$ stellar mass $M_{\ast}$ to our explicit feedback model. These sub-grid models (by construction) lower the SFR, but in both cases leave the qualitative behavior of the SFH identical. Explicit feedback models not only suppress the total $M_{\ast}$ formed, but change the shape of the SFH. SF is more ``bursty'' on small timescales, and the SFR is flatter in time (more biased to late times, without the broad high-$z$ peak).
    \label{fig:sfh.subgrid}}
\end{figure}

\begin{figure*}
    \centering
    \plotsidesize{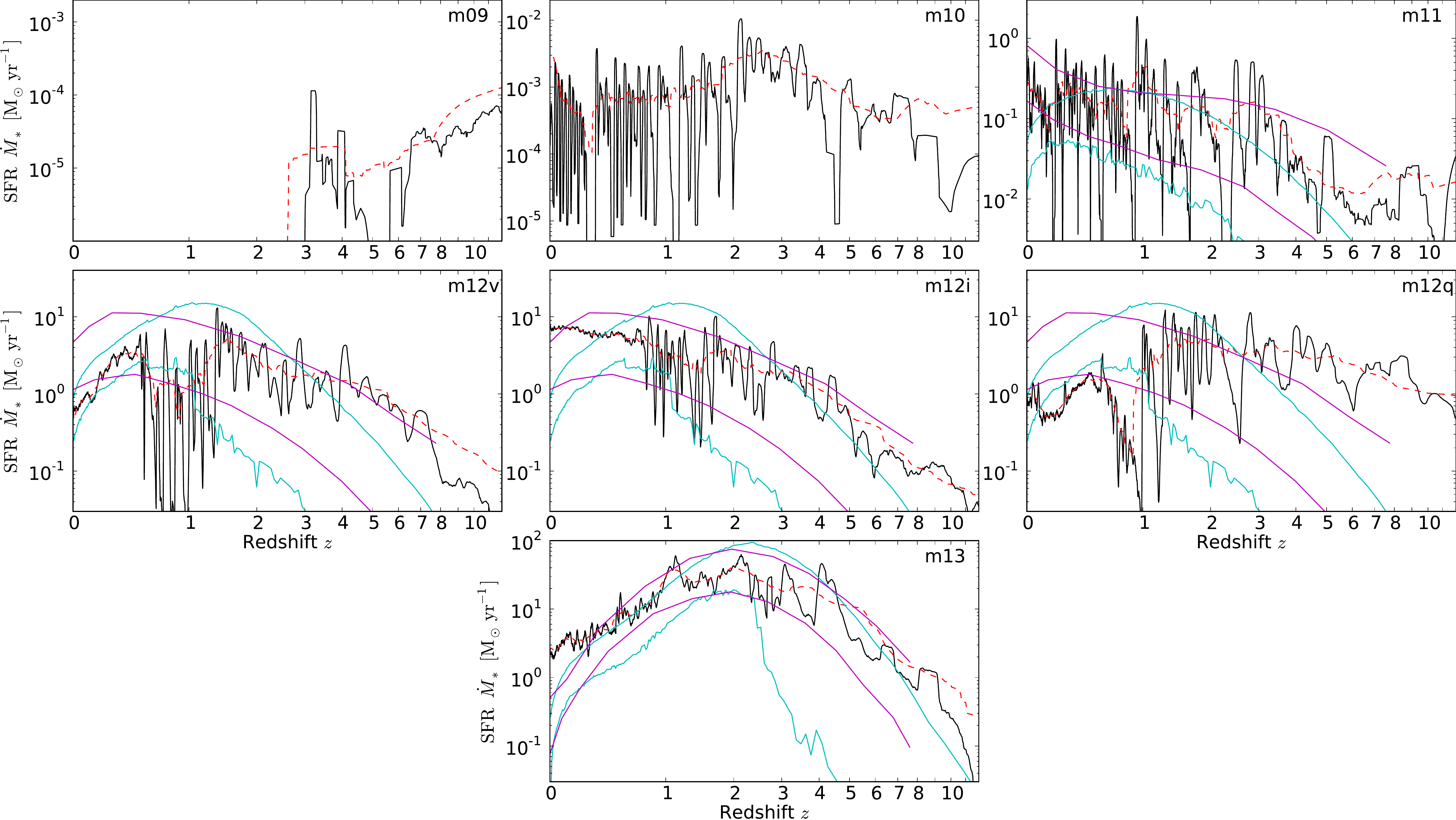}{1.01}
    \caption{SFH for each ``main'' (largest) $z=0$ galaxy in our standard (explicit-feedback) simulations. Lines show the mean SFR averaged on timescales of $10^{8}$\,yr (black solid) and $10^{9}$\,yr (red dashed). With explicit feedback, SFRs are highly variable below the galaxy dynamical time. Moreover, the (averaged) SFRs tend to be flat and/or rising with time. In contrast, with no feedback, the SFH has a sharp rise and fall peaking at $z\sim2-6$. 
    In the least massive dwarfs ({\bf m09}; $M_{\ast}<10^{6}\,\msun$ and $V_{\rm vir}(z=0)<20\,{\rm km\,s^{-1}}$), the SFR is strongly suppressed after reionization once a combination of the ionizing background and some small amount of feedback from the early star formation is able to expel most of the halo gas and prevent new cooling. 
    We compare our ${\bf m11}$, ${\bf m12}$, and ${\bf m13}$ runs to the observationally inferred ``mean'' tracks (colored lines) for the main galaxies in halos of the same $z=0$ mass, from \citet[][magenta]{moster:2013.abundance.matching.sfhs} and \citet[][cyan]{behroozi:2012.abundance.matching.sfhs}. In each case the lines bracket the $1\,\sigma$ range/scatter in the observed galaxy population. Our {\bf m11} and {\bf m12} runs agree very well with these constraints; however, in the most massive systems ({\bf m13}), the galaxy never ``quenches,'' and the SFR remains high in conflict with observations below $z\sim1$. 
    \label{fig:sfh.z0}}
\end{figure*}

\begin{figure*}
    \centering
    \hspace{-1.1cm}
    \plotsidesize{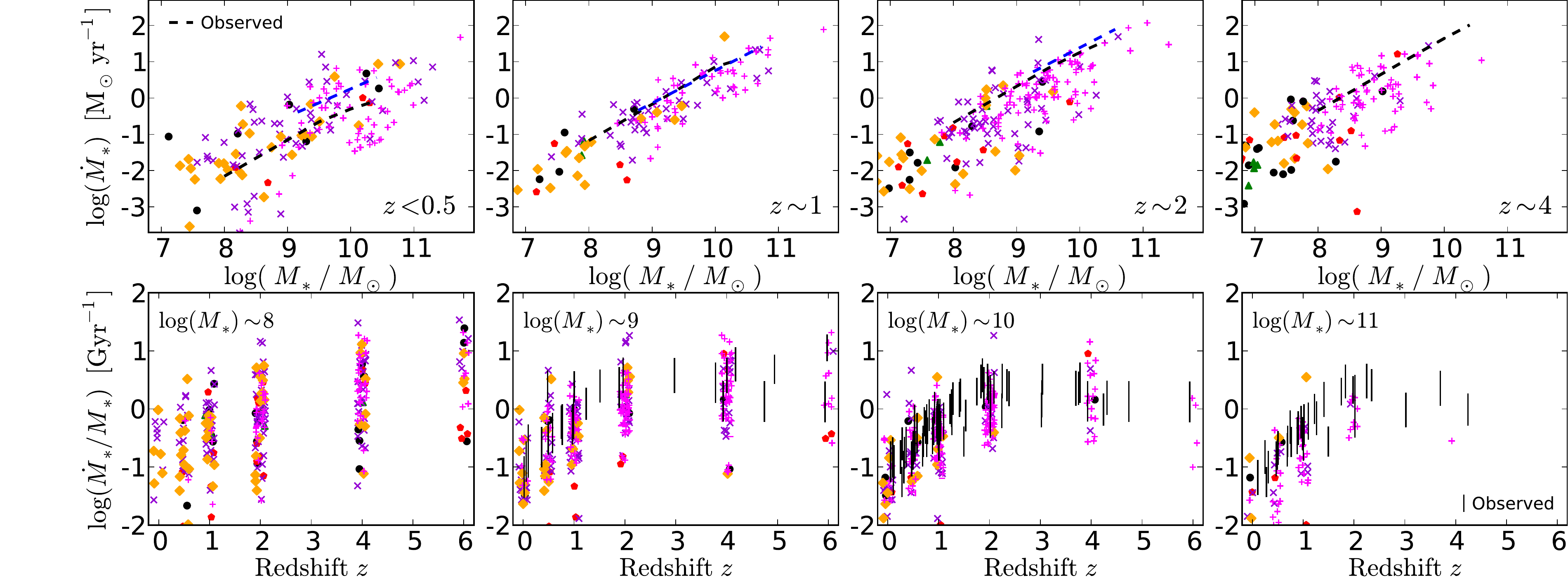}{1.04}
    \caption{{\em Top:} SFR versus galaxy stellar mass at different redshifts. We compare the observed (best-fit) relations from the compilation in \citet[][black dashed]{behroozi:2012.abundance.matching.sfhs} and \citet[][blue dashed]{zahid:2012.mass.metallicity.sfr.mass.compilations} (the systematic offset is typical of different calibrations). Allowing for the typical factor $\sim2$ systematic observational calibration uncertainty, the agreement is good at all $z$. However, magenta $+$'s compare low-resolution ($100$\,pc softening) runs of some massive halos which produce too-massive galaxies at $z=0$: there is little offset between these simulations and our fiducial models. The observed relation is simply a consequences of galaxies having relatively ``flat'' star formation histories.
    {\em Bottom:} Specific SFR of galaxies with different $M_{\ast}$, versus redshift. Observations are compiled in \citet[][Table~5]{behroozi:2012.abundance.matching.sfhs} and \citet{torrey:2013.arepo.cosmo.sim.vs.obs}. The dynamic range here is smaller so the plot appears noisier, but the information is identical to that at {\em top}. SSFRs at $z\gtrsim2$ are relatively flat, indicating rising SF histories at high-$z$. 
    \label{fig:sfrs}}
\end{figure*}

\vspace{-0.5cm}
\subsection{Instantaneous Suppression of Star Formation (at Fixed Gas Densities)}
\label{sec:results:ks}

We now examine galaxy star formation rates. In the previous section, we showed that the {\em integrated} SF is suppressed with feedback. But equally important is that feedback suppresses {\em instantaneous} SFRs in galaxies. This is manifest in the Kennicutt-Schmidt (KS) relation, shown in Fig.~\ref{fig:ks}.\footnote{We define $\Sigma_{\rm SFR} = \dot{M}_{\ast}/\pi\,R_{\rm SFR}^{2}$ (where $\dot{M}_{\ast}$ is the total SFR and $R_{\rm SFR}$ is the half-SFR radius) and $\Sigma_{\rm gas}=M_{\rm gas}/\pi\,R_{\rm SFR}^{2}$ (where $M_{\rm gas}$ is the gas mass within the $90\%$ SFR radius). Defining both $\dot{M}_{\ast}$ and $M_{\rm gas}$ within $R_{\rm SFR}$ or the stellar effective radius shifts the points along the relation.} We plot the simulations at all redshifts (the redshift evolution is insignificant), and compare to observations at a range of redshifts (which also find little or no evolution).\footnote{We compile the observed local galaxies in \citet{kennicutt98} and \citet{bigiel:2008.mol.kennicutt.on.sub.kpc.scales}, and high-redshift galaxies in \citet{genzel:2010.ks.law} and \citet{daddi:2010.ks.law.highz}; shaded region shows the $90\%$ inclusion range at each $\Sigma_{\rm gas}$ from the compilation. As discussed in those papers, there is no significant offset between the high and low-redshift systems at fixed $\Sigma_{\rm gas}$.} 

The predicted KS law agrees well with observations at all redshifts. As shown in \paperone-\paperthree, this emerges naturally as a consequence of feedback, and is {\em not} put in by hand. Recall that the instantaneous SF efficiency (SF per dynamical time) in dense gas in the simulations is $100\%$; however the global SF efficiency is $\sim2\%$. This difference arises because at $\sim2\%$ efficiency, feedback injects sufficient momentum to offset dissipation (indeed, given the same feedback, we obtain the identical KS law independent of the details of our small-scale SF law; see \citealt{hopkins:rad.pressure.sf.fb,hopkins:virial.sf}). 

If we instead consider simulations with weak/no feedback, the global KS relation is severely over-predicted (efficient cooling leads to global efficiencies $\sim100\%$). In most cosmological simulations, this is offset ``by hand'' by simply enforcing a large-scale SF law that is sufficiently ``slow'' that it agrees with the observations; however we see that this is already (implicitly) a sub-grid feedback model. Including explicit feedback obviates the need for these prescriptions, meaning that the instantaneous SF properties are truly predictive, and not simply a consequence of our chosen small-scale SF law.

\vspace{-0.5cm}
\subsection{Global Star Formation Histories}
\label{sec:results:sfh}

In Fig.~\ref{fig:sfh.subgrid}, we examine the SFH of one MW-mass galaxy. We compare this to common sub-grid models. First, a ``no-feedback'' model following \citet{springel:multiphase}; this includes only a sub-grid model for the effects of stellar feedback on the ISM structure (an ``effective equation of state'') which ensures, by design (via tuned parameters) that the galaxy lies on the Kennicutt-Schmidt relation and has reasonable gas densities. However, without galactic winds, gas from inflows quickly builds up and the SFR rises until $\dot{M}_{\ast}\approx \dot{M}_{\rm inflow}$, and nearly all the baryons are turned into stars. The galaxy at $z=0$ is far too massive, and most of its stars are old (formed at $z\gtrsim2$, with the SFR peaking at $z\sim 5$).\footnote{As noted in the many previous simulations with this sub-grid model \citep[e.g.][]{springel:lcdm.sfh}, the ``effective equation of state'' approach does not allow cooling below $\sim10^{4}\,$K, so if those simulations properly included molecular or fine-structure cooling (with the appropriate high resolution), the SFH might peak even earlier.} We then add a sub-grid wind model, in which gas is ``kicked'' out of the galaxy (forced to free-stream to ensure it escapes the disk) at a rate proportional to the SFR: here the mass-loading is equal to the SFR (``sub-grid wind 1''). This suppresses the SFR (as it is intended to do), by about a uniform factor $\sim2$, as expected. However this still leaves a too-massive galaxy, with most of its stars formed very early. Next, we consider a stronger wind model (``sub-grid wind 2''): the mass-loading is doubled, with the free-steaming length fixed. This further suppresses the SFR -- in this model the final stellar mass agrees reasonably well with our explicit-feedback simulation. However, the sub-grid model still produces a SFR which peaks at very high redshifts $z\sim2-6$. The problem is that in all the sub-grid models -- regardless of the absolute suppression of the integrated SFR or position on the Kennicutt-Schmidt relation -- the {\em shape} of the galaxy SFH still closely resembles the shape of the halo inflow rate vs.\ time \citep[for examples of this with other sub-grid models, see][]{oppenheimer:metal.enrichment.momentum.winds,scannapieco:2011.mw.mass.gal.form,stinson:2013.new.early.stellar.fb.models,puchwein:2013.by.hand.wind.model}. 

These broadly peaked SF histories are disfavored by a variety of observations. They produce too-massive galaxies at high redshift, as discussed above. But they also produce galaxies with SF histories at high-$z$ that disagree with direct observational constraints \citep[see][]{papovich:highz.sb.gal.timescales,reddy:z2.lbg.spitzer,stark:2009.lbg.sfhs}. 

With our full, explicit feedback model included, we see that the {\em shape} of the SF history is qualitatively changed, and is more consistent with observations. At all times, SFRs are much more time-variable (this is discussed below). At the highest $z\gtrsim6$, halo and stellar masses both grow efficiently (albeit with some offset).\footnote{We caution that for our massive galaxies ($M_{\rm halo}\gtrsim10^{12}\,\msun$; with particle masses $\sim10^{4}\,\msun$), at high redshifts ($z\gtrsim4$), the progenitor galaxies have small baryonic masses and so are not as well resolved. As a result, the SF histories at these masses and redshifts depend more sensitively on the details of how feedback is coupled, even though the later-time SFRs and final stellar masses are robust to these variations. See Appendix~\ref{sec:appendix:algorithms} for details.} This is the ``rapid assembly'' phase, before/during reionization, in which feedback -- while able to eject some gas from the galaxy and provide some overall suppression and variability of $\dot{M}_{\ast}$ -- does not appear to dominate the gas dynamics (the central potential and mass of the halo grow on timescales comparable to the galaxy dynamical time; so $\dot{M}_{\ast}\propto \dot{M}_{\rm halo}$). But from $z\sim2-6$, feedback acts strongly, and there appears to be a maximum, steady-state SFR which is constant or slowly increasing with time at which the galaxy is able to cycle new material into a fountain and so maintain equilibrium. This ``quasi-equilibrium'' SFR scales with the {\em central} potential of the galaxy (see \paperthree), as traced by quantities such as the central halo density or $V_{\rm max}$ (the maximum circular velocity), {not} the halo mass or virial velocity. The central potential depth increases only weakly over this time as halos accrete material on their outskirts. Below $z\sim2$, a competition ensues between slowing halo accretion rates and more highly-enriched halo gas raising cooling rates. Individual mergers also have a more dramatic effect on SF histories. 

In Fig.~\ref{fig:sfh.z0}, we show the SFH for each main $z=0$ galaxy in our simulations,\footnote{We define this as the formation rate vs.\ time (essentially a histogram of formation times) of all stars which end up in a $10\,$kpc aperture centered on the final ($z=0$) main galaxy in the simulation, averaged in $10^{8}\,$yr bins. Since most of these stars form ``in situ,'' the results are similar if we instead identify the most massive progenitor galaxy at all times and plot its galaxy-integrated SFR at each time.} and see that all cases with $10^{9}\lesssim M_{\rm halo} \lesssim 10^{13}\,\msun$ exhibit similar (relatively flat or slowly rising) SFHs.\footnote{Interestingly, the {\bf m12q} simulation shows a much higher high-$z$ SFR than {\bf m12v} or {\bf m12i}. This is in part because the particular choice of a ``quiescent'' halo led, in this case, to a halo with relatively little growth at late times ($z\lesssim3$), hence a particularly early ``formation time.''} In the most massive halos, some decline occurs when $M_{\rm halo}\gtrsim10^{12}\,\msun$, as the cooling time of virialized gas becomes longer relative to the dynamical time (the system transitions to ``hot mode'' accretion and filamentary infall is suppressed). However, we stress that the galaxies are clearly not ``quenched'' -- every system we simulate is still very much a star-forming, blue galaxy at $z\sim0$ (our {\bf m13} simulation would need a SFR $\ll 1\,\msun\,{\rm yr^{-1}}$ at $z=0$ to be ``red and dead'' by most definitions, but its SFR is $\sim5\,\msun\,{\rm yr^{-1}}$). In very low mass halos (e.g.\ our {\bf m09}, with $V_{\rm vir}(z=0)<20\,{\rm km\,s^{-1}}$), cooling is strongly suppressed after reionization.

\vspace{-0.5cm}
\subsection{Specific SFRs and the SF ``Main Sequence''}
\label{sec:results:mainsequence}

Fig.~\ref{fig:sfrs} compares the galaxy-integrated SFRs in all our simulated systems (including non-main halos) with observations of the SFR or specific SFR (SFR$/M_{\ast}$) as a function of galaxy stellar mass, at various redshifts. The simulations agree well with the SFR ``main sequence'' (SFR$-M_{\ast}$ relation) observed at all $z$ (observations plotted include compilations from 
\citealt{erb:lbg.gasmasses,noeske:2007.sfh.part1,daddi:2007.sfr.z2.exhaustion.time,elbaz:2007.sfr.mass.relation,stark:2009.lbg.sfhs} and others in \citealt{behroozi:2012.abundance.matching.sfhs} (see Table~5 therein) and \citealt{zahid:2012.mass.metallicity.sfr.mass.compilations}. The scatter is also similar to that observed. There may be some slight tension (the predictions being slightly high at $z=0$ and low at $z=2$), but these are well within the range of systematic uncertainties owing to different SFR calibrations (we show a couple such examples). By extension, the simulations similarly agree with the evolution in specific SFRs of galaxies as a function of mass. 

Evolution in specific SFRs and SFR$-M_{\ast}$ towards higher SSFR at high-$z$ simply reflects rising gas fractions (as it must, since the simulations lie on the same KS-law in Fig.~\ref{fig:ks}). The ``flattening'' of SSFR at high-$z$ implies SF histories of individual galaxies are rising with time (as we see directly); physically it follows from the saturation of gas fractions at large values, and rapid growth of halo mass at these times. The SFR$-M_{\ast}$ relation is, to lowest order, just $\dot{M}_{\ast} \sim M_{\ast}/t_{\rm Hubble}(z)$ -- this must be trivially true in any scenario where SFRs are relatively flat and/or rising with time (typical of star-forming galaxies). For this reason we see the same relation even in our simulations without feedback, as have other simulations with different feedback prescriptions \citep[see][]{keres:fb.constraints.from.cosmo.sims,dave:2011.mf.vs.z.winds}. And we see that even the very massive halos (which produce ``too large'' an $M_{\ast}$ at low redshifts) lie on the extension of the observed relation (the problem is that they continue on the relation, rather than ``quenching'' and moving below it, as observed at high masses).


\begin{figure}
    \centering
    \hspace{-0.3cm}
    \plotonesize{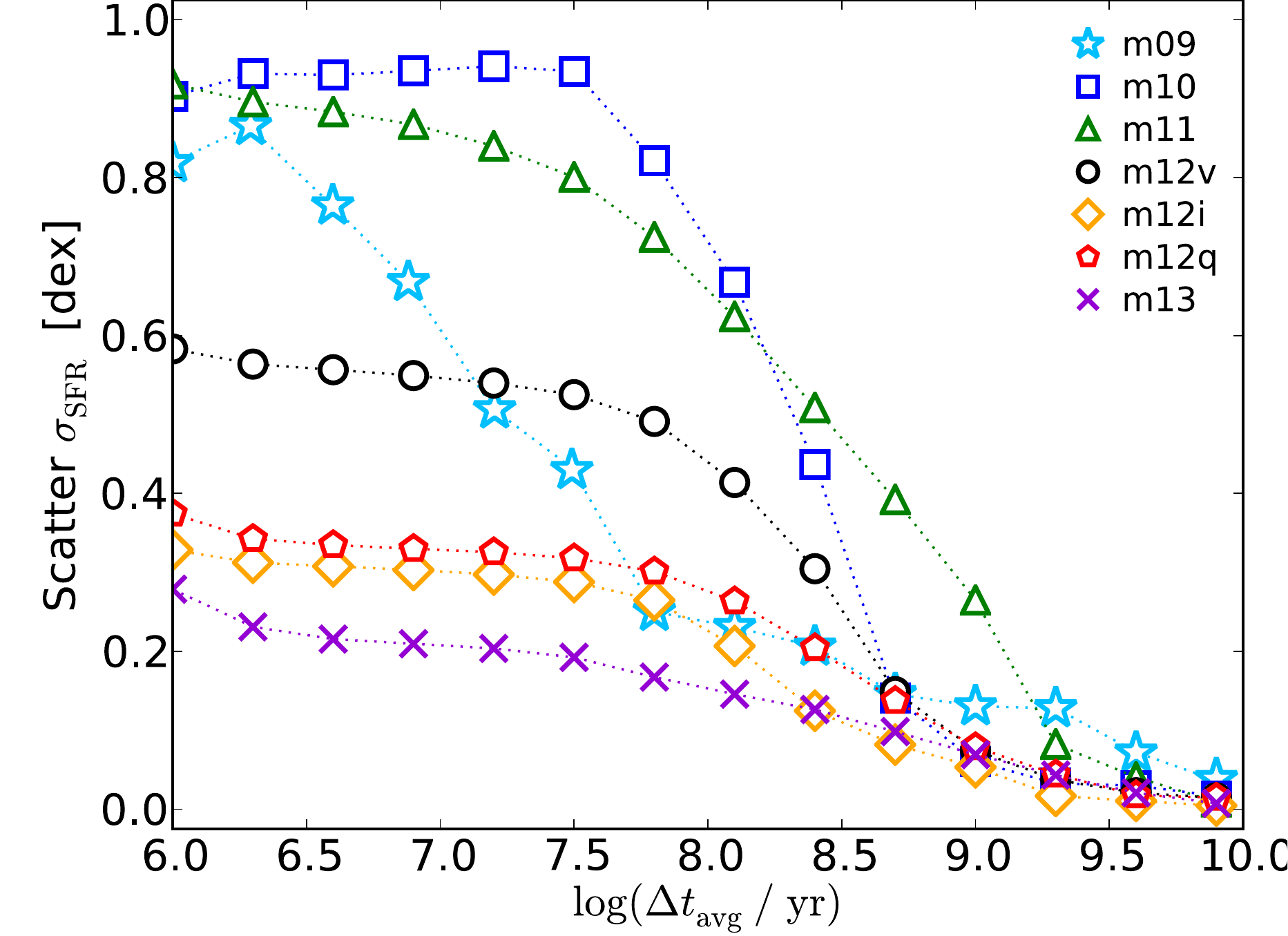}{1.02}
    \vspace{-0.2cm}
    \caption{Variability of the SFHs shown in Fig.~\ref{fig:sfh.z0}, quantified versus timescales $\Delta t_{\rm avg}$. For each ``main'' galaxy in each simulation, we show the logarithmic dispersion in the SFR $\sigma_{\rm SFR}$ about its mean on longer timescales, when the SFR is time-averaged over the timescale $\Delta t_{\rm avg}$. The variability rises substantially on timescales $\sim10^{7}-10^{8}$\,yr (galaxy dynamical times), owing to a combination of fountain dynamics, local structure in the galaxies, and stochastic effects from individual star forming regions. The short-timescale variability is a factor of $\sim2-3$ in $\sim L_{\ast}$ galaxies, but rises to order-of-magnitude level in dwarfs (where individual star clusters and bursts have a more dramatic effect).
    \label{fig:sfh.burstiness}}
\end{figure}

\vspace{-0.5cm}
\subsection{Quantifying Burstiness/Variability in SFRs}
\label{sec:results:burstiness}

In Fig.~\ref{fig:sfh.subgrid}, we showed that the SFRs are significantly more time-variable in models with explicit/resolved feedback as compared to sub-grid feedback models. We quantify this in Fig.~\ref{fig:sfh.burstiness}. We measure the dispersion in the SFR smoothed over various time intervals. Unsurprisingly, the scatter is larger on small timescales. On $\gg10^{8}\,$yr timescales, the variability is always small (SFHs are ``smooth'') -- this is more a function of the evolution of the halo over a Hubble time. Some such long-timescale variability is driven by mergers and global gravitational instabilities, but much of the short-timescale variability is not connected to these phenomena. Rather, on smaller timescales (comparable to the galaxy dynamical time) the dynamics of fountains, feedback, and individual giant molecular clouds and star clusters becomes important, so the scatter increases down to timescales $\sim10^{6}\,$yr (comparable to the massive stellar evolution timescale).\footnote{Note that even in the Milky Way, a large fraction of the observed star formation is associated with just the few most massive GMCs, so cloud-to-cloud variations can have significant effects on the global SFR \citep{murray:2010.sfe.mw.gmc}.} The short-timescale scatter is modest ($\sim0.3\,$dex) for massive systems ($M_{\rm halo}\gtrsim10^{12}\,\msun$), but rises in smaller halos (where even single star clusters can have large effects) to $\sim1\,$dex at $M_{\rm halo}\lesssim 10^{10}\,\msun$.\footnote{We have studied this in our resolution tests and found it is relative robust to spatial resolution, though the variability {\em increases} artificially on small timescales if the mass resolution is poor (factor $\sim10-100$ larger particle masses than we use), since single star particles then represent very massive star clusters.}

\vspace{-0.5cm}
\section{Discussion \&\ Conclusions}
\label{sec:discussion}

\subsection{Key Results and Predictions}


We present a series of cosmological zoom-in simulations$^{\ref{foot:movie}}$ of galaxies with $M_{\rm halo}\sim10^{9}-10^{13}\,\msun$ and $M_{\ast}\sim10^{4}-10^{11}\,\msun$. At this time, several of these runs represent the highest-resolution in both mass and force resolution of any fully cosmological runs to $z=0$. But the most important improvement, compared to previous simulations, is that we for the first time include a fully explicit treatment of both the multi-phase (cold molecular through atomic, ionized, and hot diffuse) ISM and stellar feedback. Our treatment of the ISM is enabled both by our resolution and improved treatment of cooling and heating physics (e.g.\ molecular and metal line cooling, photo-ionization and photo-electric heating with self-shielding). Our stellar feedback model utilizes explicit time-dependent energy, momentum, mass and metal fluxes taken {\em directly} from stellar population models, without free/adjustable parameters. As such, the SFRs in our simulations, the resulting outflows, and galaxy stellar masses are {\em not} the result of tuning or ``by hand'' adjusting feedback efficiencies. Critically, we include not just thermal energy from SNe, but the momentum and energy associated with SNe Types Ia \&\ II, stellar winds (young star \&\ AGB), local photo-ionization and photo-electric heating, and radiation pressure from UV and IR photons. In addition, our formulation of SPH resolves the historical numerical problems with this method, especially important for cooling in hot halo atmospheres (see Appendix~\ref{sec:appendix:sims}; \citealt{hopkins:lagrangian.pressure.sph}; Keres et al.\ in prep). 

\, \ \ \\
Our key conclusions include:

\begin{itemize}

\item{Stellar feedback -- from known sources including SNe (energy {\em and} momentum), stellar winds, radiation pressure (primarily optical/UV), and photo-heating -- is both necessary and sufficient to explain the observed relation between galaxy stellar mass and halo mass, and by extension the shape of the galaxy mass function and clustering, at stellar masses $M_{\ast}\lesssim 10^{11}\,\msun$. This appears to be true at all redshifts.}

\item{No {\em one} feedback mechanism alone is sufficient: the effects add non-linearly, and the common approximation in simulations of including only SNe feedback severely over-predicts galaxy masses (especially at low masses and/or high redshifts). The effects are even worse if the feedback momentum is ignored (if only thermal energy is considered).}

\item{The $M_{\ast}-M_{\rm halo}$ relation evolves very weakly with redshift (because outflow efficiencies depend mostly on the {central} binding energy within the galaxy). At $z\gtrsim2$, weak evolution towards higher $M_{\ast}(M_{\rm halo})$ at low masses is equivalent to a steepening faint-end slope of the galaxy luminosity function, similar to what is inferred observationally \citep{bouwens:highz.sfh,stark:2009.lbg.sfhs}.}

\item{Stellar feedback and standard cooling physics explain low galaxy stellar masses, but do not appear sufficient to explain ``quenching'' (late time suppression of star formation in massive halos) -- none of our massive systems are ``red and dead.''}

\item{Our simulations reproduce the observed Kennicutt-Schmidt relation. This is despite the fact that we assume a small-scale SF efficiency of $100\%$ in self-gravitating dense gas. As such, the KS law and instantaneous SFRs are truly predicted, not simply a consequence of our sub-grid SF law. The low star formation efficiency we find is a consequence of stellar feedback, {\em not} the microphysics of how stars form in dense gas. Absent feedback, efficient cooling leads to a global SF efficiency of $\sim100\%$ per dynamical time. With feedback -- from the same mechanisms that produce large-scale outflows and regulate galaxy formation -- the SF efficiency self-regulates at $\sim2\%$, the level where feedback injects sufficient momentum to offset dissipation.}

\item{Realistic feedback changes the {\em shape} of galaxy star formation histories. In particular, feedback from stellar radiation (both photo-heating and radiation pressure) is critical for disrupting dense, cold gas, and so is especially important for suppressing star formation in high redshift galaxies. This leads to much flatter, or gently rising, star formation histories in sub-$L_{\ast}$ galaxies. Most previous sub-grid models give qualitatively different results, in conflict with observations.}

\item{The observed star formation ``main sequence'' and specific SFRs emerge naturally from the shape of the galaxies' star formation histories (from $M_{\ast}\sim10^{8}-10^{11}$ and $z\sim0-6$). This includes ``flat'' SSFR evolution at $z\sim2-6$.
However these are relatively insensitive to feedback, since any broadly flat or rising SF history predicts $M_{\ast}(z)\sim t_{\rm Hubble}(z)\,\langle \dot{M}_{\ast}(z)\rangle$, consistent with the observations.}

\item{Dwarf galaxies exhibit much more ``bursty'' SF histories, with large variability in their SFRs on short timescales ($\sim1\,$dex scatter on $\lesssim10^{7}\,$yr timescales). This is because star formation and star cluster formation, and their associated feedback, are stochastic. The variability is not driven by mergers or global gravitational instabilities. Massive ($\sim L_{\ast}$) galaxies are much less variable ($\sim0.3\,$dex scatter in SFRs). This may translate into significantly larger scatter in $M_{\ast}(M_{\rm halo})$ at dwarf masses compared to $\sim L_{\ast}$ galaxies.}

\end{itemize}

\vspace{-0.5cm}
\subsection{Numerical Methods}

We see relatively weak dependence on simulation resolution, which is perhaps surprising given the small-scale structure present in the ISM. However, in Papers {\small I-III} \&\ Appendix~\ref{sec:appendix:test.isolated}, we presented extensive resolution studies of isolated disk galaxies simulated using the same prescriptions but numerical resolution varied from values comparable to those here, to order-of-magnitude superior mass and spatial resolution. We showed that the galaxy-averaged SFR is one of the very first quantities to converge, and is consistent to within factor $\sim2$ even for relatively poor resolution: this is because it traces the {\em integral} effect of feedback balancing turbulent dissipation. That said, we {\em do} see qualitative changes in behavior if the resolution falls below that needed to meaningfully resolve ISM phase structure, at which point ``self-regulation'' by feedback loses meaning. As a rule of thumb, the simulations must at least resolve the Toomre/Jeans length and mass (the size of the largest GMCs) in each star-forming disk, and the results are especially numerically stable if the mass resolution can be ideally pushed to $< 10^{4}\,\msun$. Even in this regime, quantities such as the phase structure of dense gas and outflows are much more sensitive to resolution, and will be discussed in more detail in future work. 

Given the same feedback model, we also see little difference in the stellar mass buildup between our standard simulations, run with a numerical algorithm designed and shown to eliminate essentially all major differences between grid (Eulerian) and smoothed-particle (Lagrangian) hydrodynamics methods, and an older version of SPH that exhibits large differences in test problems. Thus we expect little or no difference between the results here and those from adaptive-grid or moving-mesh codes, if the same feedback and ISM physics could be included. This owes to two key points: first, the differences between numerical methods in this respect, even where significant, are generally much smaller than the orders-of-magnitude differences owing to the inclusion or exclusion of the relevant physics. The stellar mass content of galaxies is set by the total amount of feedback injected, and so it is unsurprising that the time-averaged star formation rate is insensitive to changes in the detailed phase structure of the gas around galaxies. Second, the numerical differences primarily affect mixing instabilities in multi-phase, sub-sonic, pressure-dominated gas. As such, many comparison studies have shown that while the numerical differences can be important for details of the structure of hot halos in massive galaxies, they are generally unimportant inside cold star-forming gas, or in sub-$L_{\ast}$ halos, where the flows of interest tend to be highly super-sonic and gravity-dominated \citep[see e.g.][]{kitsionas:2009.grid.sph.compare.turbulence,price:2010.grid.sph.compare.turbulence,bauer:2011.sph.vs.arepo.shocks,sijacki:2011.gadget.arepo.hydro.tests}. As one considers more detailed galactic properties, we expect the differences between numerical methods to manifest as discrepancies in the cooling properties, phase structure, or distribution of heavy elements in the CGM, and to impact the way in which both inflowing cool gas and feedback-driven outflows interact with gas in galactic halos. For these reasons, an accurate numerical scheme is critical if one hopes to study the detailed structure of both gas in and around galaxies with realistic feedback. A much more extensive comparison of numerical methods is presented in a companion paper \citep{keres:2013.fire.cosmo.vs.numerics}.

\vspace{-0.5cm}
\subsection{Future Work}

This is a first exploration of cosmological simulations with explicit stellar feedback models, and many open questions remain. We have studied the effects of realistic stellar feedback on galaxy star formation histories and stellar masses; however, a complete understanding of this self-regulation requires a much more detailed examination of the dynamics of galactic outflows. In companion papers, we will study how outflows are generated, and how these interact with the circum-galactic and inter-galactic medium. It will be particularly important to build new observational diagnostics and explore whether or not different feedback mechanisms lead to different observable properties in the ISM, CGM, and IGM \citep{faucher-giguere:2014.fire.neutral.hydrogen.absorption}. Complementary questions regarding the morphology of galaxies -- how the sizes, bulge-to-disk ratios, kinematics, and other properties of the simulated systems here depend on different feedback mechanisms -- will be developed as well. The resolution and explicit treatment of the ISM in these simulations make possible many additional studies. 

Going forward, it will also be important to examine the role of additional physics. Some other physics is probably needed to explain the ``quenching'' of star formation in massive systems ($M_{\rm halo}\gg10^{12}\,\msun$). AGN feedback is a plausible candidate, which we have studied in previous work using idealized sub-grid models for the ISM. But the consequences could easily be completely different in a resolved multi-phase medium. Other physics such as magnetic fields, anisotropic conduction, and cosmic rays may be important as well, and their consequences are just beginning to be explored \citep[e.g.][]{jubelgas:2008.cosmic.ray.outflows,hanasz:2013.cosmic.ray.winds,salem:2013.cosmic.ray.outflows}.

\vspace{-0.7cm}
\acknowledgments 
We thank the many friends and peers who discussed this work in progress and sent helpful suggestions after the first draft was posted to the arXiv.
The simulations here used computational resources granted by the Extreme Science and Engineering Discovery Environment (XSEDE), which is supported by National Science Foundation grant number OCI-1053575; specifically allocations TG-AST120025 (PI Keres), TG-AST130039 (PI Hopkins), and TG-TG-AST090039 (PI Quataert). Collaboration between institutions for this work was partially supported by a workshop grant from UC-HiPACC.
Partial support for PFH was provided by NASA through Einstein Postdoctoral Fellowship Award Number PF1-120083 issued by the Chandra X-ray Observatory Center, which is operated by the Smithsonian Astrophysical Observatory for and on behalf of the NASA under contract NAS8-03060, and by the Gordon and Betty Moore Foundation through Grant \#776 to the Caltech Moore Center for Theoretical Cosmology and Physics. 
JO also thanks the financial support of the Fulbright/MICINN Program and NASA Grant NNX09AG01G. 
DK acknowledges support from the Hellman Fellowship fund at the UC San Diego and NASA ATP grant NNX11AI97G.
CAFG is supported by a fellowship from the Miller Institute for Basic Research in Science and by NASA through Einstein Postdoctoral Fellowship Award number PF3-140106 and grant number 10-ATP10-0187. 
EQ is supported in part by NASA ATP Grant 12-ATP12-0183, a Simons Investigator award from the Simons Foundation, the David and Lucile Packard Foundation, and the Thomas Alison Schneider Chair in Physics.
\\

\vspace{-0.2cm}
\bibliography{/Users/phopkins/Documents/work/papers/ms}

\begin{appendix}

\section{Baryonic Physics: Details of the Algorithmic Implementation}\label{sec:appendix:algorithms}

\subsection{Cooling}

Gas cooling is solved implicitly each timestep (using the iterative algorithm from {\small GADGET-3}). Heating/cooling rates are computed including free-free, photo-ionization/recombination, Compton, photo-electric, metal-line, molecular, and fine-structure processes. We follow $11$ separately-tracked species (H, He, C, N, O, Ne, Mg, Si, S, Ca, and Fe, each with its own yield tables associated directly with the different mass return mechanisms below; see \citealt{wiersma:2009.enrichment}). The appropriate ionization states and cooling rates are tabulated from a compilation of {\small CLOUDY} runs, including the effect of a uniform but redshift-dependent photo-ionizing background computed in \citet{faucher-giguere:2009.ion.background}, together with local sources of photo-ionizing and photo-electric heating (described below with the relevant feedback mechanisms). Self-shielding is accounted for with a local Sobolev/Jeans-length approximation (integrating the local density at a given particle out to a Jeans length to determine a surface density $\Sigma$, then attenuating the background seen at that point by $\exp{(-\kappa_{\nu}\,\Sigma)}$). Confirmation of the accuracy of this approximation in radiative transfer experiments can be found in \citet{faucher-giguere:2010.lya.cooling.selfshield} and \citet{rahmati:2013.selfshield.rt}. With this accounting, metal-line cooling follows the rate tables from \citet{wiersma:2009.coolingtables}, free-free rates follow \citet{katz:treesph}, and photo-electric rates follow \citet{wolfire:1995.neutral.ism.phases}. Compton heating/cooling is included both from the CMB and local sources, accounting as in \citet{cafg:2012.egy.cons.bal.winds} for possible two-temperature plasma effects at very high temperatures by limiting the Compton rates by the Coulomb energy exchange rates (though in practice this is only relevant at much higher temperatures than are seen in these simulations). Fine-structure and molecular cooling at low temperatures ($T<10^{5}\,$K) is tracked using an interpolation table for a compilation of {\small CLOUDY} runs as a function of the density, temperature, metallicity, and local ionizing background, as in \citet{robertson:2008.molecular.sflaw}. A temperature floor is included at the maximum of either $10\,$K or the CMB temperature at the given redshift. 

\vspace{-0.5cm}\subsection{Star Formation}

Star formation occurs probabilistically. At each timestep ${\rm d}t$, a gas particle has a probability of turning into a star particle $p = 1-\exp{(-\dot{m}_{\ast}^{i}\,{\rm d}t/m^{i}_{\rm gas})}$, where $\dot{m}_{\ast}^{i}$ is the SFR integrated over the particle, and $m_{\rm gas}^{i}$ is the particle gas mass. The SFR is non-zero {\em only} for particles with density above $n>n_{\rm crit}$ (generally $n_{\rm crit} = 100\,{\rm cm^{-3}}$), which are also locally self-gravitating using the criteria developed in \citet{hopkins:virial.sf} ($\alpha\equiv \delta v^{2}\,\delta r/G\,m_{\rm gas}(<\delta r) \approx \beta\,(|\nabla\times{\bf v}|^{2}+|\nabla\cdot{\bf v}|^{2})/G\,\rho < 1$, with $\beta\approx0.25$), and which have a non-zero molecular fraction $f_{\rm mol}>0$. The molecular fraction is determined following \citet{krumholz:2011.molecular.prescription}, using the local Sobolev approximation and metallicity to estimate the integrated column to dissociating radiation ($\tau \approx \kappa\,\langle \Sigma\rangle$ with $\langle \Sigma \rangle = \rho\,[h_{\rm sml} + (\nabla\,\ln{\rho})^{-1}]$ and $\kappa = \kappa_{\rm gas} + \kappa_{\rm dust,\,MW}(Z/Z_{\odot})$). The SFR per unit volume for gas that meets all of these criteria is then $100\%$ per free-fall time, $\dot{\rho}_{\ast}=\rho_{\rm mol}/t_{\rm ff} = f_{\rm mol}\,\rho_{\rm gas}/t_{\rm ff}$. When a gas particle becomes a star particle, the star particle inherits the metallicity of each followed species from its parent, and the conversion/formation time of the particle is used to determine its age in subsequent timesteps. The star particles also inherit their mass and gravitational softening from their parent gas particles, so that they are not over or under-resolved relative to the medium from which they form.\footnote{As discussed in the main text, we have confirmed (as seen in previous simulations of isolated galaxies in \citealt{hopkins:rad.pressure.sf.fb} and \citealt{hopkins:virial.sf}) that with feedback active, the star formation prescription makes little difference to our results. We have re-run our {\bf m09}, {\bf m10}, and {\bf m12v} runs removing the virial criterion and/or molecular criterion from the star formation law, changing the SF density threshold from $\sim10-1000\,{\rm cm^{-3}}$, and changing the SFR per free-fall time from $\sim10\%-200\%$. These changes yield only small (factor $<2$), random (non-systematic) changes to the star formation history and resulting stellar mass.}

\vspace{-0.5cm}\subsection{Stellar Feedback}

The stellar feedback algorithms follow those developed in \paperone-\papertwo, with some modifications necessary for cosmological simulations.

{\bf Radiation Pressure:} Gas surrounding stars (see below) receives a direct momentum flux $\dot{P}_{\rm rad} \approx (1-\exp{(-\tau_{\rm UV/optical})})\,(1+\tau_{\rm IR})\,L_{\rm incident}/c$ where $1+\tau_{\rm IR} = 1+\Sigma_{\rm gas}\,\kappa_{\rm IR}$ accounts for the absorption of the initial UV/optical flux and multiple scatterings of the re-emitted IR flux if the region between star and gas particle is optically thick in the IR (assuming the opacities scale linearly with gas metallicity). At each timestep we evaluate the optical depth in a smoothing kernel around the star particle (whose pre-absorption stellar spectrum $L_{\nu}$ is tabulated as a function of age and metallicity). The UV/optical absorption ($\tau_{\rm UV/optical} = \int\,\kappa\,\rho\,{\rm d}\ell$) is estimated via the Sobolev approximation (as above) in multiple frequency bins (see \papertwo\ for details). The absorbed fraction of $L$ is then distributed within the SPH smoothing kernel according to each particles' relative contribution to the optical depth. This absorbed luminosity is assumed to re-radiate isotropically in the IR (and while it can be re-absorbed, it is again re-radiated in the IR), so for an (assumed) gray-body opacity it imparts an acceleration $a_{\rm IR} = \kappa_{\rm IR}\,F_{\rm IR}/c$ to all gas in the kernel.\footnote{If the gas is optically thick in the IR out to the edge of the smoothing kernel, the kernel is iteratively expanded so this region is treated explicitly. But this is almost never the case at our resolution.} 

Photons which are not absorbed in the UV/optical, and the re-emitted IR flux, define an effective ``emergent spectrum'' for each kernel. This is propagated to large distances in the gravity tree, where it is used to calculate the local incident flux on all gas particles from stars outside the smoothing kernel; the same frequency-dependent opacities are used to calculate the local absorption and momentum flux ($\dot{P} = L_{\rm abs}/c$). The momentum flux is imparted in every timestep along the direction determined by the flux-limited diffusion approximation (along the local optical depth gradient).\footnote{Note that this avoids the ``clump detection'' algorithm described in \paperone. That was important there, where regions which were optically thick in the IR (e.g.\ cores within GMCs) could be well-resolved, so coherent radiative transfer effects required a means to estimate the clump ``membership'' and so avoid artificial numerical cancellation of momentum (which would occur if momentum fluxes were randomly oriented within the clump, as shown in Fig.~A1 therein). The importance is decreased here because such regions are always at most marginally resolved. But in either case, computing the optical depth gradient allows us to recover a very close approximation to the previous information without the added neighbor search (which is very expensive in cosmological simulations), and is more faithful to full radiative transfer in the diffusion limit. In Appendix~\ref{sec:appendix:test.isolated} we show both algorithms give very similar results in the isolated galaxy simulations from \paperone. We have also re-run our {\bf m12v} simulation with the older ``clump detection'' algorithm and find it has little systematic effect on any quantities considered here. We also note that since the radiative acceleration is implemented as a continuous force term (as recommended in \paperone), there is no need to estimate an ``escape'' or ``kick'' velocity in the calculation. And we emphasize that there is no ``boost factor'' in the equations above, only {\em resolved} absorption.}

{\bf Photo-Ionization and Photo-Electric Heating:} Here the algorithm is identical to that in \papertwo. We first tabulate the rate of production of ionizing photons for each star particle (as a function of age and metallicity); moving radially outwards from the star, we then ionize each neutral gas particle until the photon budget is exhausted (using the gas density, metallicity, and ionization state to determine the necessary photon number). Note that -- unlike the purely local Stromgren sphere approximation sometimes used in the literature -- this accounts for whether each particle is already ionized, and (if so) allows the ionized region to continue to expand (thus accounting for large coherent/overlapping HII regions).\footnote{For numerical convenience, we cut off this walk at the minimum of $\sim5$\,kpc or $50$ times the particle smoothing length. Remaining photons are assumed to have escaped.} In the cooling routine, ionized gas is flagged as having a sufficiently strong local ionization field to keep it fully ionized for the duration of the timestep; this local ionizing field information together with the escaped UV flux defined above is also used to determine the photo-electric heating rates in the cooling routine.

{\bf Supernovae and Stellar Winds:} The SNe Type-I and Type-II rates are tabulated from \citet{mannucci:2006.snIa.rates} and {\small STARBURST99}, respectively, as a function of age and metallicity for all star particles; this determines a probability per unit time ${\rm d}p = ({\rm d}N_{\rm SNe}/{\rm d}M_{\ast}\,{\rm d}t)\,m_{\ast,\,i}\,{\rm d}t$ which we use to determine whether a SNe occurs in a given particle each timestep drawing from a Poisson distribution.\footnote{It is possible, in principle, to have multiple SNe in a single particle and timestep; however, particle masses and timesteps in the dense regions where young stars live are sufficiently small that this is quite rare (occuring $<1\%$ of the time when there is a SNe in the particle). To be avoid rare but severe exceptions, we also enforce a timestep limiter of $10^{5}$\,yr for all stars $<30\,$Myr in age.} If so, the appropriate ejecta mass, metal yields (for all followed species), energy, and momentum are tabulated and directed radially from the star, and we assume the ejecta shocks within the gas in the smoothing kernel $h_{\rm sml}$ (appropriately weighted)\footnote{A particularly useful test problem is a series of SNe exploding at a constant rate from a fixed point in a thin disk (modeled as an infinite plane) with a low-density atmosphere out of plane. At high resolution, the correct solution shows the energy and momentum ``venting'' out of plane. In the case where the disk is poorly resolved (less than one smoothing length in height), if the SNe energy/momentum coupling among the neighbor particles/cells is weighted by the standard SPH smoothing kernel (effectively mass-weighted coupling), then almost all the momentum goes into the disk plane and artificially drives an expanding ring. We therefore chose to assign the weights according to the fraction of the surface area subtended by each particle as seen by the SNe; this recovers the correct behavior even in the limit of low resolution.} around the star. 

However, coupling this appropriately requires knowing whether the shock is energy or momentum conserving up to the scales we resolve.\footnote{Unlike the ultra-high resolution isolated galaxy simulations in \papertwo, we cannot be sure we always resolve the energy-conserving phase of SNe expansion in these simulations.} To estimate this, consider the cooling radii calculated in high-resolution simulations of individual blastwaves: 
$R_{\rm cool}\approx 28\,{\rm pc}\,E_{51}^{0.29}\,\langle n_{\rm cgs} \rangle^{-0.43}\,(Z/Z_{\odot}+0.01)^{-0.18}$ (where $E_{51}\approx1$ is the ejecta energy in units of $10^{51}\,{\rm erg\,s^{-1}}$; $\langle n_{\rm cgs}\rangle $ is the local density in ${\rm cm^{-3}}$; and $Z$ is the local gas metallicity; see \citealt{cioffi:1988.sne.remnant.evolution}). When $h_{\rm sml}\ll R_{\rm cool}$, the full ejecta (shocked) kinetic energy is coupled as thermal energy (with the ejecta momentum included as a momentum flux, and the ejecta mass and metals as mass/metal fluxes into the gas). Otherwise, at coupling, a fraction of the initial ejecta energy is instead converted from energy into momentum as would occur within the un-resolved cooling radius so the coupled momentum is ${\bf p} = {\bf p}_{\rm ej}\,\sqrt{[M_{\rm ejecta}+M_{\rm enc}(<R_{\rm cool})]/M_{\rm ejecta}}$, and an additional fraction of the shocked thermal energy is cooled away before being allowed to artificially do any work (according to $E_{\rm thermal,\,shocked}\propto (R/R_{\rm cool})^{-6.5}$; see \citealt{thornton98}). We stress that this is very different from assuming the SNe energy goes directly into a wind, or from turning off cooling in the gas. We are simply accounting for the possibility of an unresolved Sedov-Taylor phase and depositing the appropriate {\em momentum}, not just thermal energy, in the ambient medium in that case.

As noted in the text, at high redshifts, the progenitors of massive galaxies ($\gtrsim10^{12}\,\msun$) are not as well mass-resolved, given our particle masses of $\sim10^{4}\,\msun$. At $z\sim6$, the progenitor of a $z=0$, $10^{12}\,\msun$ halo has $M_{\rm halo}\sim10^{10}\,\msun$ and so should have $M_{\ast}\sim10^{7}\,\msun$, just $\sim1000$ particles. As such, the details of how momentum from SNe is coupled into a kernel of $\sim100$ particles can have significant effects on the entire baryonic galaxy. For example, we see significant changes to the SFR at $z\gtrsim4$ if we ``cap'' the momentum input at a modest value ${\bf p}\approx20\,{\bf p}_{\rm ej}$. However at lower redshifts (or in simulations with better mass resolution, such as our dwarf galaxies) this has no systematic effect (as shown in Fig.~\ref{fig:mg.mh.fb}, where we substantially vary both the power-law scaling of the SNe momentum above and the coupling weights within the SPH kernel, and see only small effects on the final stellar mass). Because of the scaling of remnant momentum with entrained mass, the key criterion is whether the {\em mass} resolution of the simulation is $\gg M_{\rm ejecta}$. In practice we find that this explicit accounting for the SNe remnant momentum has little effect when particle masses are $\lesssim1000\,\msun$. We have also re-run all our simulations using the alternative cooling radius estimate from \citet{chevalier:1974.spherical.sne.blastwave}: $R_{\rm cool}\approx 58\,{\rm pc}\,E_{51}^{0.32}\,\langle n_{\rm cgs} \rangle ^{-0.16}\,(P_{\rm gas}/10^{-12}\,{\rm dyn\,cm^{-2}})^{-0.2}$. This makes little difference to the stellar masses and star formation histories here (there is no significant difference in their consistency with observations); the primary effect is from dropping the metallicity dependence (leading to slightly less efficient feedback in low-metallicity, poorly-resolved regions).

Stellar winds are algorithmically nearly identical to SNe, except they occur continuously. We tabulate the wind mass, metal, energy, and momentum fluxes (as a function of stellar age and metallicity), and inject these into the neighboring gas identically to the SNe.\footnote{Purely for numerical convenience, we find it useful to limit the timesteps on which this occurs, so that a fixed fraction (say, $\sim1\%$) of the particle mass is lost per ``feedback step'' (which may be longer or shorter than the timesteps on which the star particle dynamics are evolved). Testing this we see it has no effect on our results, but is significantly less expensive computationally than invoking the feedback routine every dynamical timestep.}

In both cases, we include the relative gas-star particle velocities added to the wind/ejecta velocity centered on the star in calculating the initial ejecta momentum and energy fluxes, but this has very little effect in star-forming systems (since massive star winds and SNe ejecta are fast compared to the relative velocity of stars). This can, however, be significant in old stellar populations when AGB ejecta (with wind launching velocities $\lesssim10\,{\rm km\,s^{-1}}$) dominate the mass loss.

\vspace{-0.5cm}
\section{Simulation Numerical Details}
\label{sec:appendix:sims}

As noted in the text, these runs adopt the {\small P-SPH} formulation of TreeSPH in {\small GIZMO}, which features many improvements to SPH and has been tested in a wide range of problems (listed in the main text). We describe the most important differences between our numerical method and previous widely-used algorithms below.

\vspace{-0.5cm}
\subsection{SPH Formulation}
\label{sec:sims:sph}

The simulations use the Lagrangian ``pressure-entropy'' formulation of the SPH equations developed in \citet{hopkins:lagrangian.pressure.sph}. This formulation derives the SPH equations exactly from the particle Lagrangian and {\em manifestly} conserves momentum, energy, angular momentum, and entropy (in the absence of sources/sinks), and also eliminates the artificial ``surface tension'' error term which appears at contact discontinuities in previous (``density-energy'' or ``density-entropy'') formulations of SPH \citep[see also][]{saitoh:2012.dens.indep.sph}.\footnote{As noted in the text, the choice of ``density-entropy'' or ``pressure-entropy'' SPH makes relatively small differences to the predictions here. We have also re-run a few simulations using the ``pressure-energy'' form of SPH (in which the internal energy, rather than entropy, is the explicitly followed variable). In adiabatic flows with a constant timestep the pressure-entropy and pressure-energy forms are identical to machine accuracy; with adaptive timesteps, the error reduction in the latter formulation is slightly better (poorer) in cooling (adiabatic) steps. We confirm this makes little difference to our results.} The pressure-entropy formulation dramatically improves the behavior of fluid mixing instabilities (e.g.\ the Kelvin-Helmholtz and Rayleigh-Taylor instabilities), and eliminates most of the known differences between the results of grid and SPH methods for the problems of interest here \citep[for discussion of these in historical SPH implementations, see][]{agertz:2007.sph.grid.mixing,price:2010.grid.sph.compare.turbulence,bauer:2011.sph.vs.arepo.shocks}. For extensive numerical tests, see \citet{hopkins:lagrangian.pressure.sph}.\footnote{Note that non-Lagrangian schemes, in particular, have severe difficulties accurately propagating strong blastwaves and can lead to ``self-acceleration'' of particles in some regimes \citep[see][]{hopkins:lagrangian.pressure.sph}. The ``traditional'' method in {\small GADGET}, for example, (by which we refer to the implementation in \citealt{springel:entropy}) is Lagrangian, but adopts the ``density'' formulation of SPH, which introduces the problems with fluid mixing and contact discontinuities noted above.}

\vspace{-0.5cm}
\subsection{Artificial Viscosity \&\ Entropy}
\label{sec:sims:viscosity}

In SPH, the algorithm is inherently inviscid and some artificial viscosity is necessary to capture shocks. We adopt the ``inviscid SPH'' viscosity prescription with higher-order switches from \citet{cullen:2010.inviscid.sph}. This allows for excellent shock-capturing, while reducing the viscosity to identically zero away from shocks. A wide range of tests of this algorithm are presented therein. This viscosity treatment allows accurate treatment of sub-sonic turbulence down to Mach numbers $\lesssim0.1$ while simultaneously accurately capturing shocks with Mach numbers over $\sim10^{4}$.\footnote{We have compared (Kere\v{s} et al., in prep.) a number of simulations adopting the simpler, time-dependent  prescription from \citet{morris:1997.sph.viscosity.switch}. In most respects this gives very similar results, but gives higher viscosity in sub-sonic turbulence, and produces some ``particle noise'' (from interpenetration) in extremely strong shocks which can lead to unphysically high temperatures in particles leading the shock front.}

Similarly, SPH is inherently dissipationless, so a mechanism is also needed to generate mixing entropy in shocks. We implement this following \citet{price:2008.sph.contact.discontinuities}, using the same higher-order dissipation switch from \citet{cullen:2010.inviscid.sph}: this ensures that entropy exchange only occurs in crossing, sub resolution-scale flows with discontinuous entropies (i.e.\ prevents artificially multi-valued entropies). {For comparison, ``traditional'' {\small GADGET} \citep{springel:entropy} adopts a constant artificial viscosity following \citet{gingold.monaghan:1983.artificial.viscosity} with a \citet{balsara:1989.art.visc.switch} switch, and no entropy diffusion. This is far more dissipative, and smears out structure in sub-sonic turbulence as well as producing large (artificial) shear viscosities, which can lead to significant angular momentum transfer and eliminate structure in sub-sonic turbulence.}\footnote{\label{foot:bug}The first draft of this paper used the artificial viscosity prescription exactly from \citet{cullen:2010.inviscid.sph}, which produces slightly higher than desired viscosities in cosmological accretion shocks, and as implemented in our leapfrog time-integration scheme could lead to incorrect applications in the rare timesteps where the artificial viscosity term was larger than the cooling term. In all subsequent drafts this is fixed (all simulations were re-run) with an additional limiter similar to that suggested in \citet{hu:2014.psph.galaxy.tests}, and an explicit requirement that the viscosity only grows when $\nabla\cdot {\bf v} < 0$. This does not qualitatively or systematically change any of our conclusions, but does result in small differences in detail.}

We have also implemented the identical artificial viscosity and dissipation switches, and higher-order spatial gradient estimators, from \citet{read:2012.sph.w.dissipation.switches}. We have re-run a couple of our simulations using these methods, making our hydrodynamic solver essentially identical to that in {\small SPHS}. This produces only small differences, with higher mean artificial viscosity in turbulent regions, and lower in ``smooth'' spatially extended shocks \citep[e.g.\ large-scale structure; for a comparison showing this method gives similar results to grid codes for cosmological accretion/halo gas, see][]{power:2013.sphs.entropy.cores}.

\vspace{-0.5cm}
\subsection{Thermodynamic Evolution \&\ Timestep Criteria}
\label{sec:sims:timestep}

We employ a standard adaptive timestep algorithm and limiter. As shown in \citet{saitoh.makino:2009.timestep.limiter} and \citet{durier:2012.timestep.limiter}, in problems with high Mach number flows, adaptive timesteps (without a limiter) can lead to errors if particles with long timesteps interact suddenly mid-timestep with those on much shorter timesteps. Fortunately this is easily remedied by our timestep limiter, identical to that in \citet{durier:2012.timestep.limiter}. At all times, any active particle informs its neighbors of its timesteps and none are allowed to have a timestep $>4$ times that of a neighbor. Whenever a timestep is shortened (or energy is injected in feedback) particles are forced to return to the timestep calculation. {The limiter is not included in the ``traditional'' \citep{springel:entropy} {\small GADGET}.} 

We have confirmed the importance of this limiter in our simulations: if it is absent, a small number of particles in explosive blastwaves generate large energy conservation errors which can artificially over-heat under-dense regions of the IGM. Provided the switch is included it makes no difference if we restrict the timestep ratio between neighbors to $2,\,4,\,6$ or $8$.

\vspace{-0.5cm}
\subsection{Smoothing Kernel}
\label{sec:sims:kernel}

We adopt a quintic (fifth-order) spline kernel, with neighbor number designed to optimally resolve sound waves down to a wavelength $\approx h$, the ``core radius'' of the kernel.\footnote{This is calculated explicitly in \citealt{dehnen.aly:2012.sph.kernels}. Note that it is meaningless to speak of the resolution scale $h$ as the ``full size'' of the kernel out to some fixed number of neighbors for higher-order kernels, since this will change with the exact functional form of the kernel adopted even while the actual resolution scale is identical. The maximum kernel radius of compact support, and equivalently radius at which gravity reduces to exactly that of a point mass, is $\approx 3\,h$. For more discussion see \citet{price:2012.sph.review}.} The kernel size is adaptive (following the approximate ``fixed mass in kernel'' prescription in \citealt{springel:entropy}), enclosing $\sim64$ neighbors. This choice is the ``optimal'' spline kernel suggested by a wide range of tests in \citet{hongbin.xin:05.sph.kernels}, \citet{dehnen.aly:2012.sph.kernels}, and \citet{hopkins:lagrangian.pressure.sph}. However we have also experimented with the Wendland kernels in \citet{dehnen.aly:2012.sph.kernels} and triangular kernels in \citet{read:2010.sph.mixing.optimization} and see no significant improvements up to neighbor numbers $\sim500$. {For comparison, the traditional {\small GADGET} kernel is a cubic spline. This becomes unstable outside the range $\sim30-50$ neighbors; within this range, the ``effective resolution'' of the kernel is identical to that adopted here, but kernel errors are larger by nearly an order of magnitude.}

\vspace{-0.5cm}
\subsection{Gravity}
\label{sec:sims:gravity}

The gravity solver follows the {\small GADGET-3} hybrid tree-particle mesh (Tree-PM) method. However, we have modified this to allow for adaptive gravitational softenings\footnote{With this choice, the gravitational softening of the gas follows the SPH smoothing length (so there are always $\sim100$ particles inside the full softening extent), with the minimum softening/smoothing lengths set to the values in Table~\ref{tbl:sims} (fixed in physical units). These minima are chosen for numerical efficiency and the appropriate mass resolution, but also ensure that the ``hardest''  scatterings of baryons off dark matter particles are weaker than the numerical errors of the long-range gravitational forces (much smaller than the real rms accelerations from irregular structure in the disks). If adaptive softening is used for other particle types, an equivalent smoothing length is computed separately for each type: each dark matter particle is assigned a smoothing length enclosing $\sim100$ dark matter neighbors, and so on. Since stars (unlike gas and dark matter) are not actually volume-filling, it is not clear if it is more physically accurate to use adaptive softenings or fixed softenings (representing e.g.\ open cluster extents); in these simulations (which resolve ISM structure) we prefer the latter, but find that the choice makes very little difference in practice.} and to more accurately symmetrize the force between interacting particles with different softening, following the fully Lagrangian method in \citet{price:2007.lagrangian.adaptive.softening}; this manifestly maintains conservation of energy, momentum, and angular momentum. {In ``traditional'' {\small GADGET}, softenings are not adaptive, and pairwise interactions are simply smoothed by the larger of the two particle softenings.} 

We have also modified the softening kernel as described therein \citep[see also][]{barnes:2012.softening.is.smoothing} to represent the exact solution for the potential of the SPH smoothing kernel. With this change, the softening no longer represents non-Newtonian gravity; rather, the gravitational force is exactly Newtonian on small scales, but for particles which are not point masses but represent the extended mass distribution represented by the SPH kernel (matching the assumption made in the hydrodynamic equations).\footnote{This makes the force softening length in Table~\ref{tbl:sims} similar to a Plummer equivalent softening: the force becomes exactly that of a point mass (as opposed to an extended mass distribution) at $\approx 2.8\,\epsilon$ and deviates by $\sim10\%$ from a point mass at $\approx1.5\,\epsilon$.}

We have also tested most of our ``standard'' simulations with fixed gravitational softening lengths for the baryons equal to the minimum values in Table~\ref{tbl:sims}, and for the high-resolution dark-matter particles\footnote{In all our runs, the high-resolution region zoom-in region is sampled with high-resolution dark matter particles, whose properties are shown in Table~\ref{tbl:sims}. Outside of this region there is a hierarchy of progressively more massive dark matter particles (generally in ``steps'' of factors of $8$ in particle mass). These particles always have much larger softenings than the high-resolution particles, generally $>$\,kpc, chosen to approximately match the rms particle separation in the ``boundary'' region where these particles meet their higher-resolution counterparts.} equal to $(20,\,20,\,50,\,150,\,150,\,150,\,200)\,{\rm pc}$ for runs ({\bf m09, m10, m11, m12v, m12q, m12i, m13}). The latter corresponds very roughly to the dark matter softening kernel extent matching the mean inter-particle spacing at $\sim R_{\rm vir}/4$. Of course, with fixed softenings matched to Table~\ref{tbl:sims}, the baryons will be under-softened (i.e.\ can undergo hard scatterings) when they are very diffuse (in e.g.\ the inter-galactic medium), but we are most concerned with accuracy in the dense regions in the actual galaxy we are simulating, where smaller softenings are more accurate. In any case this produces only small differences ($\lesssim10\%$ in stellar mass at $z=0$). Re-running a couple simulations with fixed softenings but dark matter softenings equal to the smaller minimum values in Table~\ref{tbl:sims} (chosen to match the kernel extent and inter-particle spacing at $\sim10$\,kpc in each simulation) similarly produces little effect.
We have also re-run several of our simulations using identical softenings for both the baryonic and dark matter particles (with that softening taken to be about the geometric mean of the two values in Table~\ref{tbl:sims}). The differences are again small, comparable to our slightly lower-resolution runs; if we compare to a run with different baryon/dark matter softenings but the baryonic softening matched to these new runs, the differences are almost completely eliminated. There is almost no difference, between the models above, in the predicted central dark matter profiles (O\~{n}orbe et al., in prep). 

\vspace{-0.5cm}
\subsection{Domain Decomposition \&\ Parallelization}
\label{sec:sims:domain}

The simulation architecture has been heavily optimized from previous versions of the code. Gravity is still solved with a TreePM algorithm, with nested PM grids solving the large-scale forces while the tree is used for small-scale interactions. But the tree walk, domain decomposition, feedback routines, and SPH density and hydrodynamic force calculation have all been optimized substantially relative to the {\small GADGET-3} implementation (increasing the memory requirement by a factor of $\sim3-5$, but decreasing run-times and load-imbalances by a factor $\sim5$). The code has also been optimized for hybrid OpenMP+MPI application, allowing near-ideal strong scaling to $\approx256$ cores and modest gains to $512$ cores in pure-MPI mode, and positive strong scaling (a decrease in wall-clock time for runs of fixed size, albeit less than ideal) to $>1000$ cores in hybrid mode for runs presented in this work.

\vspace{-0.5cm}
\section{Testing in Idealized, High-Resolution Galaxy Simulations}
\label{sec:appendix:test.isolated}

\begin{figure}
    \centering
    \plotonesize{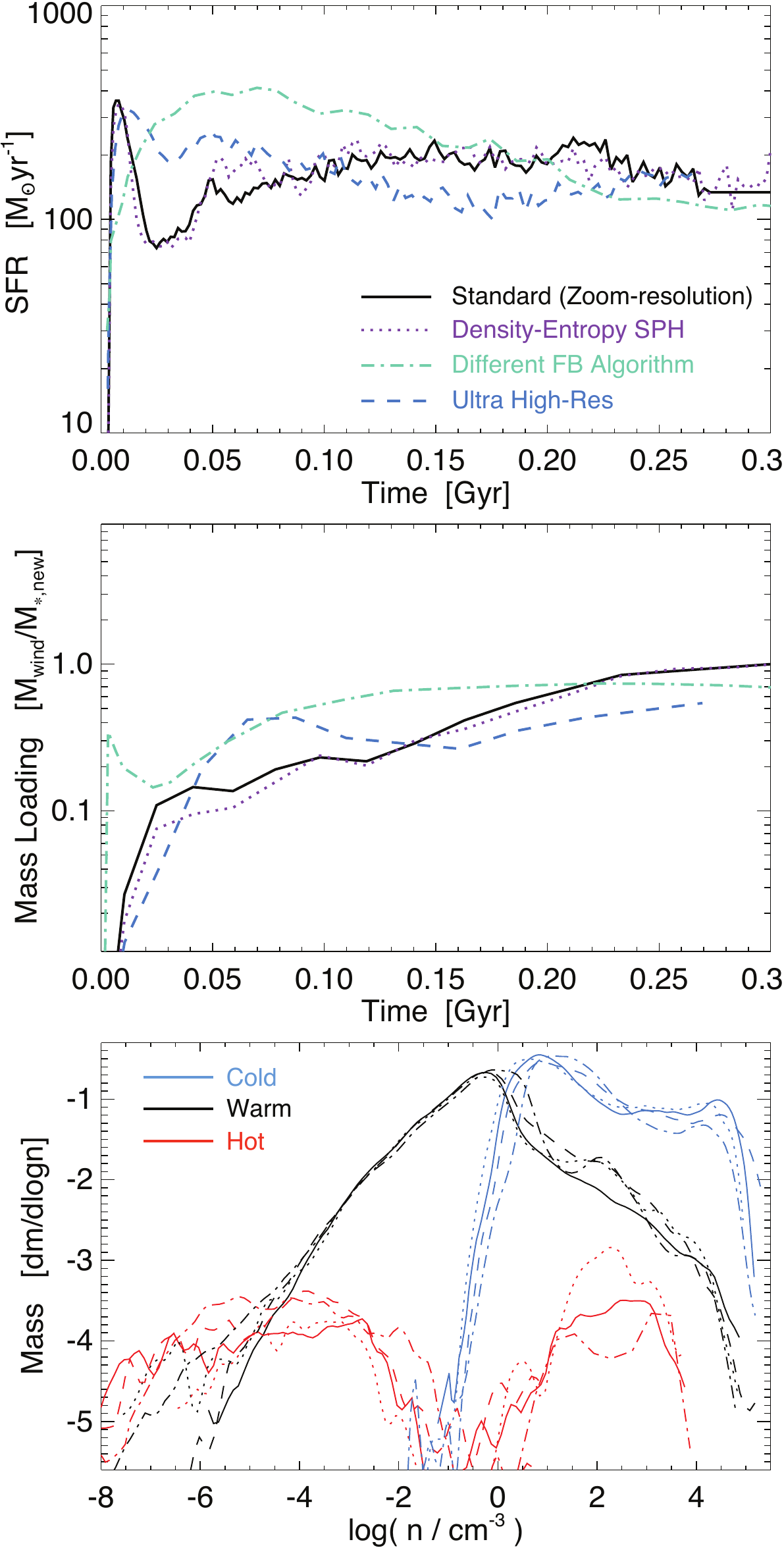}{0.98}
    \caption{Tests of our methodology in idealized simulations of a single, isolated (non-cosmological), gas-rich massive disk galaxy. {\em Top:} SFR versus time. {\em Middle:} Wind mass-loading (unbound mass $M_{\rm wind}$ versus total mass in new stars formed since the beginning of the simulation). {\em Bottom:} Distribution of gas densities at fixed time $\approx200\,$Myr after the beginning of the simulation. We separate different phases by temperature: cold gas ($T<5000\,$K), warm gas ($5000\,$K$<T<10^{5}\,$K) and hot gas ($T>10^{5}$\,K; the bimodal distribution reflects low-density, high volume-filling factor material which has escaped the disk, and hot bubbles actively heated by SNe within it). In each, we compare four runs with identical initial conditions (different linestyles, as labeled). (1) A model run with our standard numerical method, at resolution about equal to our zoom-in simulations of massive galaxies. (2) A run using the density-entropy form of SPH, with our improvements to the numerical method from \S~\ref{sec:appendix:sims} removed. (3) A run using the exact same feedback algorithms used in \papertwo-\paperthree, without the optimizations for cosmological runs described in \S~\ref{sec:appendix:algorithms}. (4) A run at the ultra-high resolution from those papers, with much better spatial and mass resolution than can be achieved in cosmological runs. We see very little difference between the runs, suggesting both that our results are stable with respect to resolution and that the changes to the code and resolution for cosmological simulations do not fundamentally alter our conclusions from the previous work.    
    \label{fig:hiz.comparison}}
\end{figure}

In a series of papers (\paperone-\paperthree), we have extensively studied and tested the feedback models used here in even higher-resolution simulations of idealized (non-cosmological) individual model galaxies. Our dwarf galaxy simulations here are run at essentially the same resolution as the ``ultra-high resolution'' dwarf galaxy models in these earlier papers, so we can safely apply the same feedback models. However, as noted in the text, in the simulations of massive halos ($\gtrsim10^{12}$) here, we are forced to lower resolution (comparable to the ``high-resolution,'' but not ``ultra-high resolution'' simulations therein); it is therefore important to check the results of the isolated galaxy simulations at lower resolution as well.

To this end, we have re-run the ``HiZ'' simulation from \paperone-\paperthree, with the identical code used for the cosmological simulations here. This was the most massive system considered therein, a disk with properties typical of star-forming galaxies at $z\sim2-4$. The halo, stellar bulge, stellar disk, and gas disk have masses $M_{\rm halo}=1.4\times10^{12}\,\msun$, $M_{\rm bulge}=0.7\times10^{10}\,\msun$, $M_{\rm \ast,\,disk}=3\times10^{10}\,\msun$, and $M_{\rm gas}=7\times10^{10}\,\msun$, with scale-lengths for the gas and stellar disk $h_{\rm \ast,\,disk}=1.6\,{\rm kpc}$ and $h_{\rm gas}=3.2\,{\rm kpc}$. We re-run this at two resolutions: first, the ``ultra-high resolution'' level used in our earlier papers, with force softening $\approx3$\,pc, and particle mass $\approx800\,h^{-1}\,\msun$. Second, at resolution about equal to our massive cosmological simulations, force softening $\approx10\,$pc and particle mass $\approx 5\times10^{4}\,\msun$.

The results are shown in Fig.~\ref{fig:hiz.comparison}. We specifically compare the SFR and wind mass-loading versus time, and the phase distribution of the gas at a time $t\approx200\,$Myr when the galaxy has reached a quasi-steady state. The mass loading is defined as the instantaneous ratio of total wind mass (defined as the gas mass which has positive Bernoulli parameter -- i.e.\ would escape in the absence of additional forces or cooling -- with outward radial velocity $>100\,{\rm km\,s^{-1}}$) to total stellar mass formed since the beginning of the simulation. In all cases, the steady-state SFR, wind mass-loading, and gas phase distribution are similar (to within a factor $<2$). The largest differences appear at very early times, and are generally artifacts of the initial conditions (in which the disk is supported by thermal pressure, before a turbulent cascade is established); these should not appear in our cosmological simulations. 

We have also re-run this second ``cosmological resolution'' simulation with the identical feedback implementation used in \papertwo\ (as opposed to the updated algorithms described above), to examine whether the improvements to the code would substantially change our previous results. We see very little difference. Since the total energy and momentum inputs are fixed to the same levels by the initial mass function, and no fundamentally new source of feedback has been introduced, this is a reassuring indication that the results are not especially sensitive to the purely numerical details of the algorithmic implementation of feedback.

As a final comparison, we re-run the simulation using the identical feedback implementation, but adopting the ``density-entropy'' form of SPH, with a constant artificial viscosity, no entropy diffusion, a cubic spline kernel, and constant (non-adaptive) gravitational softenings. This is the form of SPH which exhibits the largest and most severe differences from grid codes in test problems (especially involving fluid mixing). But consistent with our comparison in Fig.~\ref{fig:mg.mh.fb}, we see almost no difference between this simulation and our new SPH run. This is consistent also with our comparison in \citet{hopkins:lagrangian.pressure.sph}, where we perform a similar comparison with a model dwarf (SMC-mass) galaxy. At those masses we saw almost no difference in SFR, and $\sim50\%$ differences in the wind mass-loading owing to purely numerical effects, mostly the effect of fluid mixing instabilities altering the cooling time of hot SNe ``bubbles.'' In the HiZ galaxy model considered here, these differences are further minimized because the hot gas is not the dominant driver of outflows.

\end{appendix}

\end{document}